 \documentclass[final,authoryear,5p,times,twocolumn]{elsarticle}

\usepackage{amssymb}
\usepackage{color}
\definecolor{olivegreen}{rgb}{0,0.6,0}

\providecommand\aj{Astron.\ J.} % Astronomical Journal
\providecommand\aap{Astron.\ Astrophys.} % Astron and Astrophys
\providecommand\apj{Astrophys.\ J.} % Astrophysical Journal
\providecommand\apjl{Astrophys.\ J.} % Astrophysical Journal, Letters
\providecommand\icarus{Icarus} % Icarus
\newcommand\pasj{PASJ} % Publications of the ASJ
\providecommand\nat{Nature} % Nature
\providecommand\pasj{PASJ} % Publications of the Astronomical Society of Japan
\providecommand\mnras{Mon.\ Not.\ R.\ Astron.\ Soc.} % Monthly Notices of the RAS
\providecommand\planss{Planet.~Space~Sci.} % Planetary Space Science
\journal{Icarus}
\bibpunct{(}{)}{;}{a}{}{,}
\usepackage{graphicx,subfigure,longtable}

\usepackage{endfloat}

\usepackage{ulem}
\usepackage{pstricks,hyperref}

\newcommand{\degr}{\ensuremath{^\circ}}

\begin{document}

\begin{frontmatter}

\title{Olivine-dominated A-type asteroids in the Main Belt: Distribution, Abundance and Relation to Families}

\author[mit]{Francesca E. DeMeo}
\author[wei]{David Polishook}
\author[nic]{Beno{\^i}t Carry}  
\author[low]{Brian J. Burt}
\author[psi,tai]{Henry H. Hsieh}
\author[mit]{Richard P. Binzel}
\author[low]{Nicholas A. Moskovitz}
\author[mhc]{Thomas H. Burbine}

\address[mit]{Department of Earth, Atmospheric, and Planetary Sciences, Massachusetts Institute of Technology, 77 Massachusetts Avenue, Cambridge, MA 02139 USA}
\address[wei]{Department of Earth and Planetary Sciences, Weizmann Institute of Science, Rehovot 0076100, Israel}
\address[nic]{Observatoire de la C™te d'Azur, Boulevard de l'Observatoire, 06304 Nice Cedex 4, France}
\address[low]{Lowell Observatory, 1400 West Mars Hill Road, Flagstaff, AZ 86001, USA}
\address[psi]{Planetary Science Institute, 1700 E. Ft. Lowell Road, Suite 106, Tucson, AZ 85719, USA}
\address[tai]{Institute of Astronomy and Astrophysics, Academia Sinica, P.O. Box 23-141, Taipei 10617, Taiwan}
\address[mhc]{Department of Astronomy, Mount Holyoke College, South Hadley, MA 01075, USA}
%%%%%%%%%%%%%%%%%%%%%%%%%%%%%%%%

\begin{abstract}
Differentiated asteroids are rare in the main asteroid belt despite evidence for $\sim$100 distinct differentiated bodies in the meteorite record. We have sought to understand why so few main-belt asteroids differentiated and where those differentiated bodies or fragments reside. Using the Sloan Digital Sky Survey (SDSS) to search for a Òneedle in a haystackÓ we identify spectral A-type asteroid candidates, olivine-dominated asteroids that may represent mantle material of differentiated bodies. We have performed a near-infrared spectral survey with SpeX on the NASA IRTF and FIRE on the Magellan Telescope. 

We report results from having doubled the number of known A-type asteroids. We deduce a new estimate for the overall abundance and distribution of this class of olivine-dominated asteroids.   We find A-type asteroids account for less than 0.16\% of all main-belt objects larger than 2 km and estimate there are a total of $\sim$600 A-type asteroids above that size.  They are found rather evenly distributed throughout the main belt, are even detected at the distance of the Cybele region, and have no statistically significant concentration in any asteroid family.  We conclude the most likely implication is the few fragments of olivine-dominated material in the main belt did not form locally, but instead were implanted as collisional fragments of bodies that formed elsewhere.

\end{abstract}

%%%%%%%%%%%%%%%%%%%%%%%%%%%%%%%%

\end{frontmatter}

% \linenumbers

%%%%%%%%%%%%%%%%%%%%%%%%%%%%%%%%
\section{Introduction}
%%%%%%%%%%%%%%%%%%%%%%%%%%%%%%%%
%What's the problem.
%What's the background concepts and research.
%What are we doing//presenting here. (what is our question, what is our project plan, what results do we present?)

The Missing Mantle Problem, also known as the Great Dunite Shortage, is a decades-old question in planetary science \citep{Chapman1986,Bell1989} that seeks to understand the perceived shortage of olivine-dominated mantle material in the main asteroid belt, identified spectrally as A-types \citep{Bus2002b,DeMeo2009taxo}. In the classical theory of asteroid differentiation, a body would form an iron-rich core, an olivine-dominated mantle, and a pyroxene-rich basaltic crust. Over time, the clear lack of observed mantle material has had various interpretations. Earlier on, the ``battered to bits'' scenario suggested through collisionary processes these bodies had been ground down below the detection thresholds of observational surveys \citep{Burbine1996}. More modern theories posit that the lack of mantle material suggests there was no significant population of differentiated bodies in the main belt, and that differentiated planetesimals instead formed much closer to the Sun, in the terrestrial planet region.  For a review of the missing mantle problem and the evolution of proposed solutions, see \cite{DeMeo2015}. For a review of differentiation see \cite{Scheinberg2015} and \cite{Wilson2015}. For a review of the dynamical history, impact scenarios, and meteorites of differentiated bodies see \cite{Scott2015}.

Searches for basaltic (crustal) differentiated material throughout the main belt have proven successful \citep[e.g.,][]{Moskovitz2008,Solontoi2012,Leith2017}. Basaltic asteroids are easily identified spectrally in both the visible and near-infrared wavelengths by the deep and narrow one-micron and deep and broad two-micron absorption bands. The interior of a differentiated asteroid should be metal-rich. Spectrally, this composition falls within the X-type - or M-type defined by the Tholen taxonomy \citep{Tholen1984} if albedo data are available and within an intermediate range - however, these spectral classes are compositionally degenerate. Due to the much larger uncertainty in spectrally identifying metal-rich asteroids, we do not explore a search for them in this work.

An additional way to search for differentiated bodies that have been heavily or completely disrupted is to identify spectral A-type asteroids, characterized by a very wide and deep 1-micron absorption indicative of large concentrations ($>$ 80\%) of olivine. Olivine-dominated asteroids were first detected by \citet{Cruikshank1984}. Once correcting for their generally very red slopes, they are close spectral matches to brachinite or pallasite meteorites \citep[e.g.,][]{Burbine2002} and are thought to represent mantle material or core-mantle boundary material of disrupted differentiated asteroids \citep{Benedix2014}. An alternate theory for pallasites put forth by \citet{Tarduno2012} is that they result from a major impact of a molten iron core from one body with the mantle of a differentiated second body. The red slopes of A-types are expected to be due to space weathering as shown in experiments as shown by \citet{Sasaki2001,Brunetto2006}. For a detailed compositional analysis of previously known olivine-dominated asteroids, see \citet{Sanchez2014}. Not all A-types, however, are expected to be differentiated material \citep{Sunshine2007}, a point explained further in the discussion section. 

\begin{figure}
  \centering
  \includegraphics[width=\textwidth]{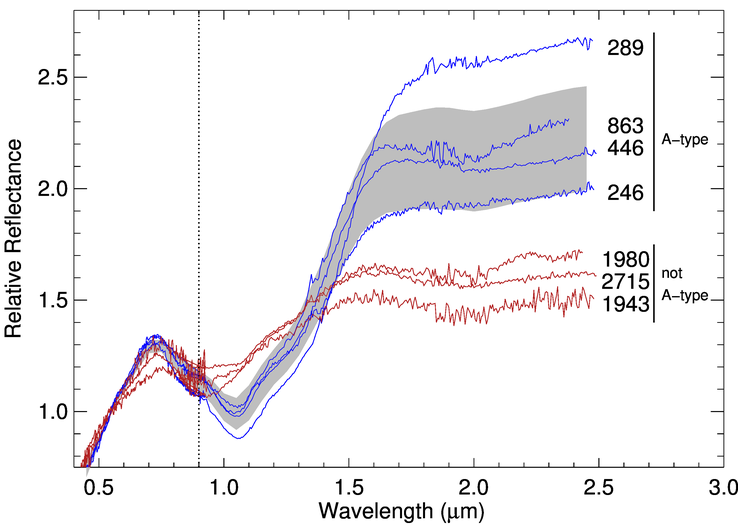}
  \caption[]{%
    	A selection of asteroid spectra that were classified as olivine-rich ÒA-typesÓ based on visible-only data \citep[from][]{Bus2002b} that diverge in the near-infrared. The dotted line at $\sim$0.9 microns marks the separation of visible-wavelength data on the left, and near-infrared data on the right. The gray shaded region marks bounds plus and minus one sigma from the mean of the A-type class defined by \citet{DeMeo2009taxo}.The blue spectra have wide and deep 1-$\mu$m bands with a band center past 1$\mu$m indicative of an olivine-rich mineralogy, while the red spectra do not, demonstrating why near-infrared data are critical to identify olivine-rich asteroids. Roughly half of all visible-data-only A-types are proven to be olivine-rich with near-infrared data \citep{DeMeo2009taxo} while the statistics are closer to 9 in 10 for basalt-rich V-types \citep{Moskovitz2008}. This figure is modified from the original version in \citet{DeMeo2007}.  }
  \label{fig: visdegen}
\end{figure}

Preliminary studies of the abundance and distribution of A-type asteroids were performed by \citet{Carvano2010} and \citet{DeMeo2013, DeMeo2014} using the Sloan Digital Sky Survey (SDSS). However, while visible wavelengths are useful to identify candidate objects, observing these potential targets in the near-infrared is critical (see Fig.~\ref{fig: visdegen}) to identify if their spectra are truly consistent with an olivine-rich mineralogy. Roughly half of all objects classified as A-types based on visible-only data do not show the characteristic strong, broad 1 micron feature in the near-ir (see Fig.~\ref{fig: visdegen}) and are thus not consistent with mantle material \citep{Burbine2002, DeMeo2009taxo}. Similarly for V-types, not all objects classified based on visible-only data are proven to be basaltic, although the success rate is roughly 90\% \citep{Moskovitz2008}.

In this work we have two goals: 1) to determine the distribution and abundance of olivine-dominated mantle material across the main belt to shed light on their origin as locally-formed or subsequently-implanted and 2) to search for differentiated fragments within asteroid families to constrain the level of differentiation in the interiors of larger parent asteroids.

For Goal 1, we identify 155 A-type candidates and take measurements of 60 of them. New observations presented here more than double the number of known A-type asteroids.  We present the spectra, determine the positive detection rate of A-type asteroids based on SDSS candidates, and calculate the total expected A-types according to their size distribution and mass. For Goal 2, we identify 69 candidates for differentiated material within families and take measurements of 33 of them, 12 of which do not overlap with the first set of observations.

\section{Observations}

  \subsection{Target Selection: General Distribution} \label{sec: targetselection}
  A major breakthrough towards efficient spectroscopic searching comes from the Sloan Digital Sky Survey (SDSS).  SDSS has been in operation for over a decade with the primary objective of imaging extragalactic objects \citep{York2000}, but also catalogs the photometric measurements (uÕ, gÕ, rÕ, iÕ, zÕ with central wavelengths 0.3551, 0.4686, 0.6166, 0.7480, and 0.8932 $\mu$m.) of moving objects that pass through its field \citep{Ivezic2001}.  Over one hundred thousand asteroids have been observed, many with multiple measurements. These measurements can be converted into spectral reflectance and can be well-characterized by two single dimensions: iÕ-zÕ, which indicates the potential depth of a 1-micron absorption band, and slope \citep[see work by ][]{Parker2008, Carvano2010, DeMeo2013}. While this low resolution and limited wavelength range certainly cannot fully characterize a body's mineralogy, it provides enough information to filter for more interesting targets. Among the hundreds of thousands of (spectroscopically) observable asteroids, finding unique or interesting objects was previously like looking for a needle in a haystack.   Searching with SDSS colors has been proven successful for basaltic asteroids \citep[e.g.,][]{Masi2008, Moskovitz2008, Duffard2009} and for identifying hydration in asteroids \citep{Rivkin2012}. SDSS is a powerful tool that allows us to focus our efforts rather than conducting blind surveys. 

\begin{figure}
  \centering
  \includegraphics[width=\textwidth]{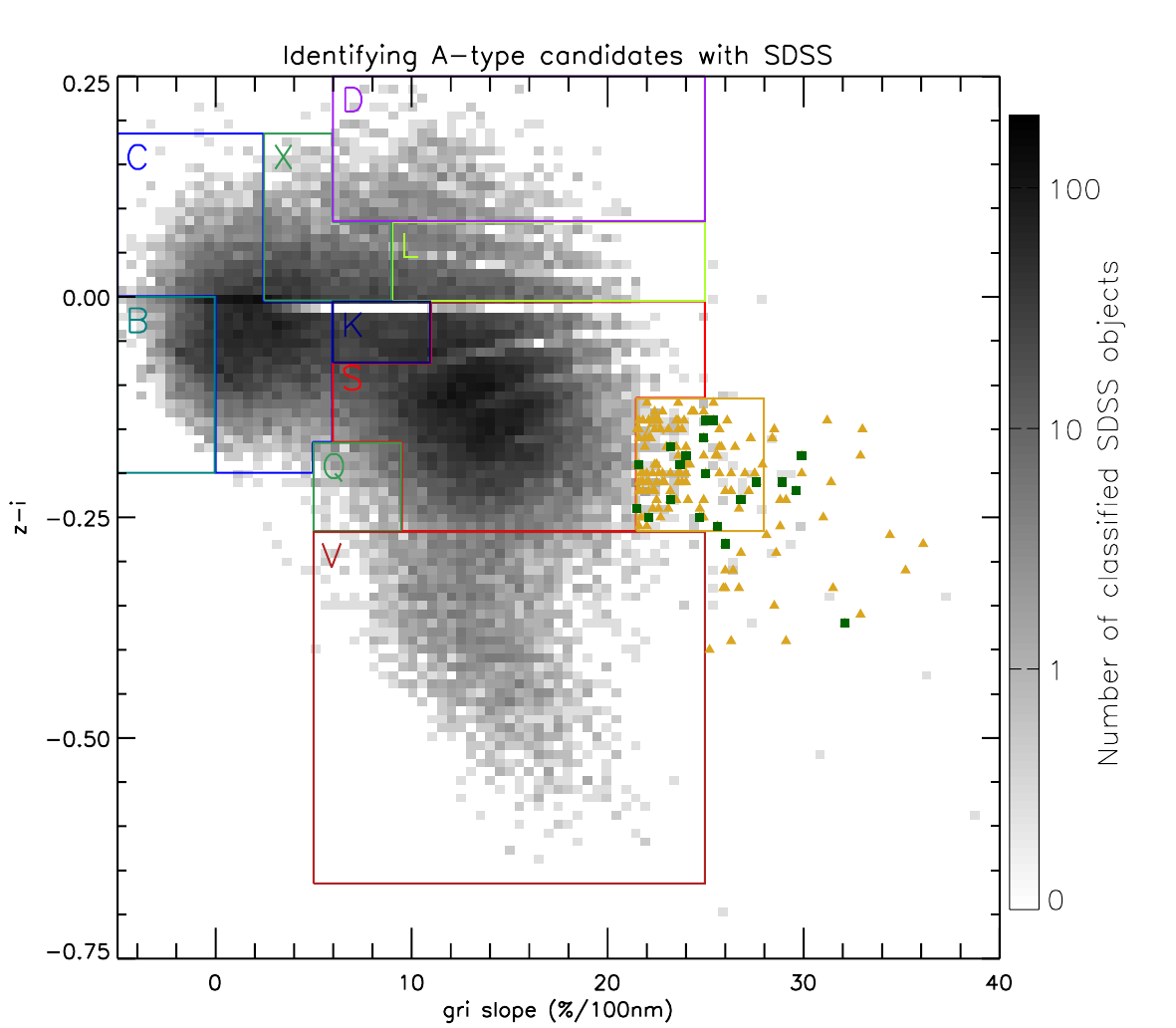}
  \caption[]{%
    	This plot shows the taxonomic boundaries for SDSS data defined in \citet{DeMeo2013}. Our A-type candidates are marked as triangles. Yellow triangles were either not observed or observed and were not A-type. Green squares are confirmed A-types. We include candidates outside the formal boundaries of A-types, because of their high slopes and indication of a 1-micron absorption band.  } 
  \label{fig: sdssbound}
\end{figure}

    Candidate A-types were chosen among objects observed in the Sloan Digital
    Sky Survey (SDSS) Moving Object Catalog (MOC). We use the fourth release
    (MOC4), including observations prior to March 2007. A subset of these data
    is selected based on quality as described in \citet{DeMeo2013}. 
    From this subset we create a list of objects with at least
    one observation with gri-slope greater than 2.15 and less then 4.0 and 
    a z-i value greater than -0.4 and less than -0.115 with class not equal to V
    (gri-slope and z-i are as defined in \citet{DeMeo2013}, see Fig~\ref{fig: sdssbound} for a visual of the
    parameter space). The area defined here is broader than for A-types in \citet{DeMeo2013}
    to increase the number of potential candidates and to explore the parameter space that 
    is not classified. We visually inspected all of the observations from this list and 
    remove objects for which the data do not look accurate (for example one point is spuriously
    high, creating the false impression of a high gri-slope). We also remove objects for which
    there are many observations, and most of which are not A-type-like.   
    Our final list includes 155 SDSS main-belt candidate A-types, 65 of which have been
    observed multiple times. The full list is provided in Supplementary Table 1 and plots of the 
    SDSS colors for these candidates are provided in Supplementary Figure 1.     
    The observational circumstances for the 60 asteroids measured spectroscopically in this work are provided in Table~\ref{tab:obs}.

    For the purposes of this statistical study we use the A-type boundary limit described above. However, we took additional observations of A-type candidates outside of those bounds to test the boundary robustness. We measured 18 A-type candidate spectra outside the formal boundaries and confirmed one as an A-type Section~\ref{sec: cybele}. See Supplementary Table 2 and Supplementary Figures 2 and 3. 
  
%%    SLOPE AS A PERCENT? NEED TO TO MULTIPLY to MATCH FIG?

%;good=where(classified.sl ge 2.15 and classified.sl lt 4.0 and classified.zi gt -0.4 and classified.zi lt -0.115 and classified.class ne 'V  ')

     \subsection{Target Selection: Families}
 Using the same SDSS data and classifications as in the previous section, we searched for A-type candidates and V-type candidates within asteroid families. We used the memberships from \citet{Nesvorny2010} to identify strict family membership, although we did include observations of candidates that were slightly outside of these boundaries, marked in the notes column of Supplementary Table 3. We identified 69 candidates for differentiation in or near asteroid families based on either being an olivine-rich (A-type) candidate or a basaltic candidate identified by an indication of a deeper than typical 1-micron absorption band (classified as V, R, Q, or U). The U class stands for `unusual` or `unclassifiable` and fell outside the boundaries of our formal classification system but are still good candidates for this work. We observed 33 candidates, 12 of which are unique to the family search, the others overlap with the full A-type distribution search described in the previous section. Most targets were in the Flora and Eunomia families. See Table~\ref{tab:obsfam} for observational circumstances.

  \subsection{SpeX Observations and Reduction}
    Observations were taken on the 3-meter NASA Infrared Telescope Facility at
    the Mauna Kea Observatory. We use the instrument SpeX \citep{Rayner2003},
    a near-infrared spectrograph in low resolution mode over 0.8 to 2.5
    $\mu$m.

    Objects are observed near the meridian (usually $<$ 1.3 airmass) in two
    different positions (typically denoted A and B) on a 0.8 x 15 arcsecond
    slit aligned north-south. Exposure times are typically 120 seconds, and we
    measure 8 to 12 A-B pairs for each object. Solar analog stars are
    observed at similar airmass throughout the night. We use the same set of
    solar analogs as the SMASS program \citep{Binzel2004,Binzel2006} that have been in use for
    over a decade. Uncertainties in spectral slope on the IRTF using these 
    consistent set of stars at low airmass is estimated to be around 5\% of the measured slope value.
    Observations were taken in good weather conditions and observations of 
    other objects throughout the night provide confidence that there were no 
    major systematic slope issues.

    Reduction and extraction is performed using the Image Reduction and
    Analysis Facility (IRAF) provided by the National Optical Astronomy
    Observatories (NOAO) \citep{Tody1993}. Correction in regions with strong
    telluric absorption is performed in IDL using an atmospheric transmission
    (ATRAN) model by \citet{Lord1992}. The final spectrum for each object is
    created by dividing the telluric-corrected asteroid spectrum by the
    average of the telluric-corrected solar star spectra throughout that
    night. More detailed information on the observing and reduction procedures
    can be found in \citet{Rivkin2004} and \citet{DeMeo2008}.

  \subsection{FIRE Observations and Reduction}
    Observations were taken on the 6.5-meter Magellan Telescope at Las
    Campanas Observatory. We use the instrument Folded-port InfraRed
    Echellette \citep[FIRE;][]{Simcoe2013} in high-throughput, low-resolution
    prism mode with a slit width of 0.8 arcsecond oriented toward the parallactic angle.  
    Exposures of 180 seconds were used for asteroids to avoid saturation due to thermal
    emission from the instrument and telescope at the long wavelength end (past 2.2 $\mu$m).

    The readout mode sample-up-the-ramp was used for asteroid observations
    requiring exposure times in multiples of 10.7 seconds. For stars readout
    mode Fowler 2 was used. Standard stars chosen were a combination of
    well-established solar analogs used for the past decade in our IRTF
    program and newly measured G2V stars that are dimmer and better suited for
    a larger, southern hemisphere telescope.  Standard stars typically needed
    to be defocused to avoid saturation. Neon Argon lamp spectra were taken
    for wavelength calibration. Quartz lamp dome flats were taken for flat
    field corrections. Observations and reduction procedures are similarly described in \citet{DeMeo2014D}

   For FIRE data reduction, we used an IDL pipeline designed for the instrument
   based on the Spextool pipeline \citep{Cushing2004}. Typically, sky correction is performed by AB pair 
   subtraction of images. In this case, because this slit is long (50'') we did not use an AB dither pattern
   and instead use nearby sky along this list from the same exposure. The FIRE reduction pipeline is built
   for this sky subtraction method.   

\begin{table*}[t]
\begin{minipage}[t]{\textwidth}
\caption{Observational Circumstances and Taxonomic Results for the Survey of A-type Candidates}
\label{tab:obs}
\begin{center}
\begin{tabular}{llllllllllll}
  \hline
Asteroid   & Designation & Telescope	& Date  & Phase & V & H & Albedo & Albedo & \# SDSS & Class & Class \\
Number   & or Name	      &	 	& (UT)	& Angle (deg)		&	Mag& Mag	& &Ref\footnote{1: \citet{Usui2011} \url{http://darts.jaxa.jp/astro/akari/catalogue/AcuA.html} 2: \citet{Masiero2011} \url{http://cdsarc.u-strasbg.fr/viz-bin/Cat?J/ApJ/741/68} 3: \citet{Masiero2012} \url{http://cdsarc.u-strasbg.fr/viz-bin/Cat?J/ApJ/759/L8} 4: \citet{Nugent2015} }	&	Obs.	& (SDSS)	& (This Survey) 	\\ 
\hline		
																			
1709	&	Ukraina	&	IRTF	&	2011/02/07	&	17.9	&	17	&	12.8	&	0.123$\pm 0.006  $	&	1	&	2	&	A,U	&	S	\\
2036	&	Sheragul	&	IRTF	&	2011/01/06	&	24.9	&	17.2	&	12.7	&	0.300$\pm 0.044  $	&	1	&	4	&	A,U,V,S	&	S	\\
2234	&	Schmadel	&	IRTF	&	2012/01/23	&	20.6	&	17.2	&	12.2	&	0.284$\pm 0.016  $	&	2	&	1	&	A	&	R	\\
3385	&	Bronnina	&	IRTF	&	2012/09/16	&	11.5	&	15.3	&	12.3	&	0.363$\pm 0.081  $	&	2	&	2	&	A,S	&	S	\\
3573	&	Holmberg	&	IRTF	&	2012/08/14	&	3	&	15.8	&	12.9	&	0.343$\pm 0.028  $	&	3	&	2	&	A,S	&	S	\\
6067	&	1990 QR11	&	IRTF	&	2012/03/20	&	18.7	&	17.3	&	11.5	&	0.256$\pm 0.060  $	&	2	&	1	&	A	&	S	\\
7057	&	1990 QL2	&	IRTF	&	2011/12/01	&	20.3	&	16.9	&	13.6	&	0.41$\pm 0.15  $	&	4	&	1	&	A	&	S	\\
7172	&	Multatuli	&	IRTF	&	2012/07/21	&	7.4	&	17.9	&	13.9	&	0.353$\pm 0.035  $	&	3	&	1	&	A	&	S	\\
8838	&	1989 UW2	&	IRTF	&	2011/12/01	&	14.1	&	16.6	&	11.4	&	0.239$\pm 0.023  $	&	2	&	9	&	9A	&	A	\\
9404	&	1994 UQ11	&	IRTF	&	2012/09/17	&	9.6	&	17.4	&	13.4	&	0.300$\pm 0.061  $	&	2	&	1	&	A	&	S/L	\\
10715	&	Nagler	&	IRTF	&	2013/07/19	&	31	&	17.2	&	13.4	&	0.309$\pm 0.040  $	&	2	&	2	&	A,A	&	A	\\
10977	&	Mathlener	&	IRTF	&	2012/03/20	&	1.2	&	17.1	&	14.9	&	0.327$\pm 0.092  $	&	2	&	1	&	A	&	A	\\
11589	&	1994 WG	&	Magellan	&	2011/07/24	&	5.1	&	17.4	&	12.5	&		&		&	4	&	4A	&	S	\\
11952	&	1994 AM3	&	IRTF	&	2012/09/17	&	1.9	&	17	&	14.1	&	0.210$\pm 0.051  $	&	2	&	2	&	S,A	&	S	\\
13577	&	Ukawa	&	IRTF	&	2012/07/22	&	6.9	&	17.4	&	14.8	&		&		&	1	&	A	&	S	\\
13724	&	Schwehm	&	IRTF	&	2012/08/15	&	1.4	&	16.4	&	14.1	&	0.474$\pm 0.233  $	&	2	&	2	&	S,A	&	S	\\
13816	&	Stulpner	&	Magellan	&	2012/03/28	&	13.3	&	18.1	&	14.2	&		&		&	3	&	2A,S	&	S	\\
16520	&	1990 WO3	&	IRTF	&	2012/01/23	&	2.7	&	17	&	14.2	&	0.325$\pm 0.178  $	&	2	&	4	&	3A,S	&	A	\\
17375	&	1981 EJ4	&	Magellan	&	2013/02/10	&	12.9	&	18.9	&	13.8	&	0.382$\pm 0.102$	&	2	&	3	&	2S,A	&	S	\\
17716	&	1997 WW43	&	IRTF	&	2013/07/19	&	4.3	&	17.4	&	15.1	&	0.434$\pm 0.142 $	&	2	&	1	&	A	&	S	\\
17818	&	1998 FE118	&	IRTF	&	2012/03/20	&	18.4	&	17.4	&	12.9	&	0.217$\pm 0.040  $	&	2	&	1	&	A	&	A	\\
17889	&	Liechty	&	IRTF	&	2013/01/17	&	8.1	&	17.5	&	14	&	0.481$\pm 0.190  $	&	2	&	1	&	A	&	A	\\
18853	&	1999 RO92	&	IRTF	&	2013/05/17	&	7.7	&	18.6	&	13.6	&	0.302$\pm 0.049  $	&	2	&	1	&	A	&	A	\\
19312	&	1996 VR7	&	IRTF	&	2013/03/01	&	10.7	&	18.6	&	14.8	&		&		&	1	&	A	&	S	\\
19570	&	Jessedougl	&	IRTF	&	2011/12/01	&	10.8	&	17	&	13.2	&		&		&	1	&	A	&	S	\\
19652	&	Saris	&	IRTF	&	2012/04/29	&	2.8	&	17.6	&	14	&	0.242$\pm 0.010  $	&	2	&	1	&	A	&	A	\\
21646	&	Joshuaturn	&	IRTF	&	2012/07/22	&	6.7	&	18.4	&	14.4	&	0.289$\pm 0.128  $	&	2	&	1	&	A	&	X	\\
21809	&	1999 TG19	&	IRTF	&	2012/09/14	&	2.2	&	17.5	&	14	&	0.168$\pm 0.036  $	&	2	&	1	&	A	&	A	\\
24673	&	1989 SB1	&	IRTF	&	2012/07/22	&	10	&	17.3	&	13.9	&	0.355$\pm 0.048  $	&	3	&	1	&	A	&	S	\\
27791	&	Masaru	&	IRTF	&	2011/12/01	&	15.8	&	17	&	14	&	0.565$\pm 0.095  $	&	2	&	1	&	A	&	S	\\
29731	&	1999 BY2	&	IRTF	&	2011/12/01	&	11.2	&	17.4	&	14.1	&		&		&	1	&	A	&	S	\\
31393	&	1998 YG8	&	IRTF	&	2012/01/23	&	2.8	&	17.3	&	14.8	&	0.467$\pm 0.116  $	&	2	&	1	&	A	&	A	\\
33745	&	1999 NW61	&	IRTF	&	2011/04/30	&	12.3	&	16.8	&	13.9	&		&		&	1	&	A	&	S/Q	\\
34969	&	4108 T-2	&	Magellan	&	2013/02/10	&	23.9	&	20.6	&	15.7	&		&		&	1	&	A	&	A	\\
35102	&	1991 RT	&	IRTF	&	2012/09/16	&	9.9	&	16.9	&	14.6	&		&		&	2	&	S,A	&	S	\\
35925	&	1999 JP104	&	IRTF	&	2011/08/08	&	4	&	17.1	&	13.9	&	0.176$\pm 0.020  $	&	2	&	1	&	A	&	A	\\
36256	&	1999 XT17	&	IRTF	&	2012/03/20	&	10.8	&	17.3	&	12.5	&	0.186$\pm 0.033  $	&	3	&	1	&	A	&	A	\\
39236	&	2000 YX56	&	Magellan	&	2013/02/12	&	17.8	&	19.9	&	15.3	&	0.102$\pm 0.017  $	&	2	&	1	&	A	&	C	\\
40573	&	1999 RE130	&	IRTF	&	2012/01/23	&	12.9	&	17.9	&	13.5	&	0.319$\pm 0.033  $	&	2	&	1	&	A	&	S	\\
44028	&	1998 BD1	&	IRTF	&	2013/07/19	&	18.9	&	18.5	&	13.3	&		&		&	1	&	A	&	S	\\
52228	&	Protos	&	Magellan	&	2012/07/11	&	20.7	&	19.1	&	13.6	&	0.271$\pm 0.176  $	&	2	&	1	&	A	&	A	\\
59115	&	1998 XG3	&	IRTF	&	2013/03/01	&	26.6	&	19.1	&	15.1	&		&		&	1	&	A	&	S	\\
60631	&	2000 FC26	&	IRTF	&	2013/03/01	&	7.5	&	18.7	&	14.5	&	0.218$\pm 0.074  $	&	2	&	1	&	A	&	A	\\
75810	&	2000 AX244	&	IRTF	&	2012/03/20	&	5	&	17.9	&	14.8	&	0.275$\pm 0.067  $	&	2	&	2	&	A	&	A	\\
77653	&	2001 KH72	&	IRTF	&	2013/07/18	&	6.2	&	18.2	&	13.9	&		&		&	1	&	A	&	S	\\
81850	&	2000 KL60	&	IRTF	&	2013/07/19	&	19.1	&	17.5	&	14.9	&	0.217$\pm 0.048  $	&	2	&	1	&	A	&	S	\\
87812	&	2000 SL146	&	Magellan	&	2011/07/24	&	21.7	&	18.9	&	14.4	&		&		&	2	&	A,X	&	X	\\
91450	&	1999 RV24	&	IRTF	&	2012/09/16	&	13.2	&	17.1	&	15.7	&		&		&	2	&	S,A	&	S	\\
92516	&	2000 ND25	&	IRTF	&	2012/03/20	&	4.7	&	17.6	&	14.8	&		&		&	2	&	S,A	&	A	\\
95560	&	2002 EX98	&	IRTF	&	2014/01/11	&	17.9	&	18.7	&	14.2	&	0.294$\pm 0.075  $	&	2	&	1	&	U	&	A	\\
97335	&	1999 YF	&	IRTF	&	2013/01/17	&	8.4	&	17.7	&	16.5	&		&		&	1	&	A	&	S	\\
98082	&	2000 RJ67	&	Magellan	&	2013/02/09	&	14	&	19.9	&	15.5	&		&		&	1	&	A	&	X	\\
105840	&	2000 SK155	&	IRTF	&	2011/12/01	&	3.7	&	17.7	&	14.2	&		&		&	2	&	A	&	A	\\
108209	&	2001 HS28	&	Magellan	&	2012/07/11	&	18.5	&	19	&	16.1	&		&		&	1	&	A	&	S	\\
111186	&	2001 WA8	&	IRTF	&	2013/07/19	&	15.3	&	18.8	&	15.4	&		&		&	1	&	A	&	C	\\
125739	&	2001 XH116	&	Magellan	&	2013/02/11	&	17.3	&	20.3	&	15.7	&		&		&	1	&	U	&	S	\\
139045	&	2001 EQ9	&	Magellan	&	2012/07/12	&	15.9	&	18.4	&	15.9	&		&		&	1	&	A	&	A	\\
175158	&	2005 EM66	&	IRTF	&	2013/07/19	&	11.3	&	18.3	&	16.5	&		&		&	1	&	A	&	S	\\
188330	&	2003 OU8	&	Magellan	&	2012/03/31	&	5.3	&	19.3	&	13.8	&	0.284$\pm 0.086  $	&	2	&	1	&	A	&	S	\\
200832	&	2001 XC238	&	Magellan	&	2013/02/09	&	10.2	&	18.5	&	15.8	&		&		&	1	&	A	&	S	\\\hline
\end{tabular}
\end{center}
\end{minipage}
\end{table*}

\begin{table*}[t]
\begin{minipage}[t]{\textwidth}
\caption{Observational Circumstances and Taxonomic Results for the Survey of Family Candidates}
\label{tab:obsfam}
\renewcommand{\footnoterule}{}
\begin{center}
\begin{tabular}{llllllllllll}
  \hline
Family & Asteroid   & Designation & Telescope	& Date  & Phase & V & H & \# SDSS & Class\footnote{We include a broad array of classes (A, R, V, Q, and U which represents "unusual" or "unclassifiable") with deeper than average absorption bands that could be differentiated candidates.} & Class & Ref\footnote{In this family survey, some candidates overlap with the A-type distribution survey. We note here if they overlap or are only part of the family survey.}\\
& Number   & or Name	      &	 	& (UT)	& Angle (deg)		&	Mag	& Mag		&	Obs.	& (SDSS)	& (This Survey) &	\\ 
\hline		
Eos	&	1297	&	Quadea	&	Magellan	&	2/10/13	&	2.9	&	15.1	&	10.9	&	4	&	A,V,S,C	&	C,X	&	family-only	\\
Flora	&	2036	&	Sheragul	&	IRTF	&	1/6/11	&	24.9	&	17.2	&	12.7	&	4	&	A,U,V,S	&	S	&	Adistrib	\\
Flora	&	7057	&	1990 QL2	&	IRTF	&	12/1/11	&	20.3	&	16.9	&	13.6	&	1	&	A	&	S	&	Adistrib	\\
Nysa/Po	&	7172	&	Multatuli	&	IRTF	&	7/21/12	&	7.4	&	17.9	&	13.9	&	1	&	A	&	S	&	Adistrib	\\
Gefion	&	7302	&	1993 CQ	&	IRTF	&	3/4/12	&	4.2	&	16.5	&	12.3	&	1	&	V	&	S	&	family-only	\\
Eunomia	&	13137	&	1994 UT1	&	IRTF	&	12/12/12	&	10	&	16.4	&	13.2	&	1	&	V	&	S	&	family-only	\\
Nysa/Po	&	13577	&	Ukawa	&	IRTF	&	7/22/12	&	6.9	&	17.4	&	14.8	&	1	&	A	&	S	&	Adistrib	\\
Eunomia	&	13816	&	Stulpner	&	Magellan	&	3/28/12	&	13.3	&	18.1	&	14.2	&	3	&	2U,S	&	S	&	Adistrib	\\
Flora	&	16520	&	1990 WO3	&	IRTF	&	1/23/12	&	2.7	&	17	&	14.2	&	4	&	3A,S	&	A	&	Adistrib	\\
Gefion	&	17818	&	1998 FE118	&	IRTF	&	3/20/12	&	18.4	&	17.4	&	12.9	&	1	&	A	&	A	&	Adistrib	\\
Vesta	&	19570	&	Jessedouglas	&	IRTF	&	12/1/11	&	10.8	&	17	&	13.2	&	1	&	A	&	S	&	Adistrib	\\
Eunomia	&	20596	&	1999 RX188	&	IRTF	&	1/11/14	&	23.8	&	19	&	14.4	&	2	&	U,V	&	S	&	family-only	\\
Eunomia	&	21646	&	Joshuaturner	&	IRTF	&	7/22/12	&	6.7	&	18.4	&	14.4	&	1	&	A	&	C,X	&	Adistrib	\\
Flora	&	24673	&	1989 SB1	&	IRTF	&	7/22/12	&	10	&	17.3	&	13.9	&	1	&	A	&	S	&	Adistrib	\\
Flora	&	27791	&	Masaru	&	IRTF	&	12/1/11	&	15.8	&	17	&	14	&	1	&	A	&	S	&	Adistrib	\\
Flora	&	28218	&	1998 YA6	&	IRTF	&	12/1/11	&	10.9	&	17.8	&	14.9	&	1	&	U	&	S,K	&	family-only	\\
Eunomia	&	30366	&	2000 JC57	&	Magellan	&	2/9/13	&	24.3	&	17.6	&	13	&	1	&	V	&	S	&	family-only	\\
Eunomia	&	33745	&	1999 NW61	&	IRTF	&	4/30/11	&	12.3	&	16.8	&	13.9	&	1	&	A	&	S/Q	&	Adistrib	\\
Flora	&	34969	&	4108 T-2	&	Magellan	&	2/10/13	&	23.9	&	20.6	&	15.7	&	1	&	A	&	A	&	Adistrib	\\
Nysa/Po	&	39236	&	2000 YX56	&	Magellan	&	2/12/13	&	17.8	&	19.9	&	15.3	&	1	&	A	&	C	&	Adistrib	\\
Eunomia	&	44028	&	1998 BD1	&	IRTF	&	7/19/13	&	18.9	&	18.5	&	13.3	&	1	&	A	&	S	&	Adistrib	\\
Flora	&	45876	&	2000 WD27	&	IRTF	&	7/22/12	&	24.1	&	18.4	&	14.5	&	1	&	U	&	S	&	family-only	\\
Eunomia	&	46456	&	4140 P-L	&	IRTF	&	7/19/13	&	23.7	&	18.1	&	14.2	&	8	&	3S2AKL,X	&	S	&	family-only	\\
Eunomia	&	55550	&	2001 XW70	&	Magellan	&	3/28/12	&	25.6	&	19	&	14.8	&	1	&	V	&	S	&	family-only	\\
Eunomia	&	66905	&	1999 VC160	&	Magellan	&	3/28/12	&	24	&	19.7	&	14.9	&	1	&	V	&	V	&	family-only	\\
Vesta	&	92516	&	2000 ND25	&	IRTF	&	3/20/12	&	4.7	&	17.6	&	14.8	&	2	&	S,A	&	A	&	Adistrib	\\
Flora	&	98082	&	2000 RJ67	&	Magellan	&	2/9/13	&	14	&	19.9	&	15.5	&	1	&	A	&	X	&	Adistrib	\\
Flora	&	108209	&	2001 HS28	&	Magellan	&	7/11/12	&	18.5	&	19	&	16.1	&	1	&	A	&	S	&	Adistrib	\\
Eunomia	&	111140	&	2001 VV96	&	Magellan	&	3/28/12	&	14.5	&	19.7	&	15.2	&	1	&	V	&	S	&	family-only	\\
Flora	&	125739	&	2001 XH116	&	Magellan	&	2/11/13	&	17.3	&	20.3	&	15.7	&	1	&	U	&	S	&	Adistrib	\\
Flora	&	139045	&	2001 EQ9	&	Magellan	&	7/12/12	&	15.9	&	18.4	&	15.9	&	1	&	A	&	A	&	Adistrib	\\
Vesta	&	200832	&	2001 XC238	&	Magellan	&	2/9/13	&	10.2	&	18.5	&	15.8	&	1	&	A	&	S	&	Adistrib	\\
Eunomia	&	222511	&	2001 TX47	&	Magellan	&	3/28/12	&	17.6	&	20.7	&	15.8	&	2	&	R,V	&	S	&	family-only	\\

\hline
\end{tabular}
\end{center}
\end{minipage}
\end{table*}

\section{Results}

\subsection{A-type spectral results, abundance and distribution} \label{sec: aresult}
From three decades of asteroid spectral observations only $\sim$15 A-type asteroids have been discovered, that have been confirmed to be olivine-dominated from near-infrared spectroscopic measurements \citep[e.g.][]{Cruikshank1984,Tholen1984,Bus2002b,deleon2004,DeMeo2009taxo,Sanchez2014,Borisov2017,Polishook2017}. In our survey we have detected 21 A-type asteroids (20 in the statistical work presented here, plus asteroid (11616) observed outside of the strict candidate boundaries) more than doubling the number of known A-types. Spectra of the confirmed A-types are plotted in Fig.~\ref{fig: specA}. The survey spectra that are not A-types are plotted in Supplementary Figure 4. We note that one asteroid (1709) is extremely red with a deep 1-micron band that is typically characteristic of an A-type, however, the band center is shifted shortward of 1.0 micron. We overplot it with an average A-type spectrum to highlight the spectral difference. We prefer to keep our sample of A-types restrictive, thus we exclude (1709) from an A-type classification. In Fig.~\ref{fig: orbits} we plot the survey results in orbital space. The orbits of the confirmed and rejected candidates as well as the unobserved candidates are shown. 

   \begin{figure}
  \centering
  \includegraphics[width=\textwidth]{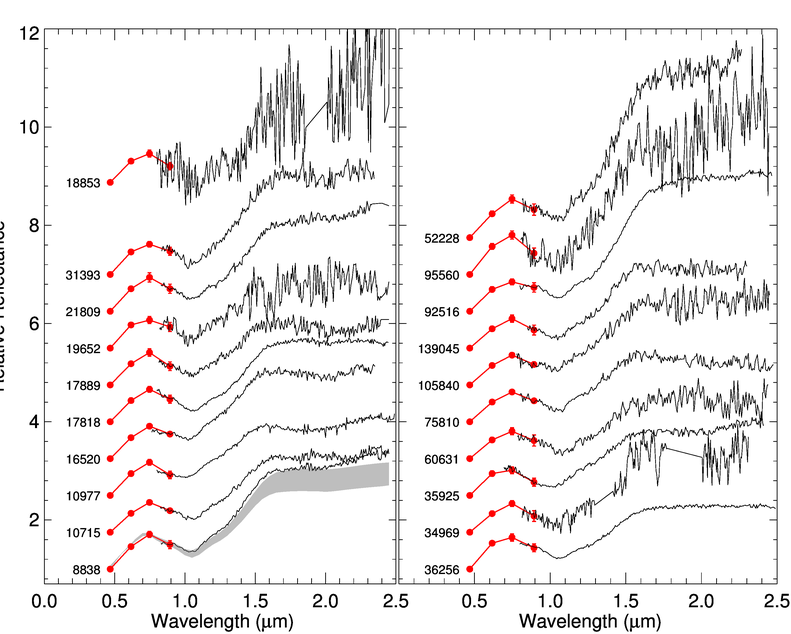}
  \caption[]{%
    	Plot of confirmed A-types based on near-infrared spectral measurements from this work. SDSS colors are plotted as red dots with red error bars. The spectra are plotted in black. The gray region, overplotted on one spectrum, bounds plus and minus one sigma from the mean of the A-type class defined by \citet{DeMeo2009taxo}. Asteroids (16520), (17818), (34569), (92516), and (139045) are dynamically associated with asteroid families and are discussed further in Sec.~\ref{sec: fam}.} 
  \label{fig: specA}
\end{figure}

We use the results of this survey to calculate the expected number of A-types throughout regions of the main belt. Overall, we confirm 20 A-types out of 60 candidates, a 33\% success rate. In fact, the success rate is the same for objects observed once and more than once by SDSS (14/42 for singly-observed and 6/18 for multiply observed objects), whereas for similar work for D-types \citep{DeMeo2014D} the success rate for multiply-observed objects was higher (1/7 or 14\% of singly-observed versus 2/5 or 40\% for multiply-observed D-type candidates). Given this success rate, for the full SDSS sample we would expect 52 A-types (155 candidates multiplied by a 33\% success rate). There are 32,453 objects in the SDSS sample (defined as having classifications in \citet{DeMeo2013} with semi-major axes 1.8 $\leq$ a $<$ 3.7 AU and H magnitude 11 $\leq$ H $<$ 17), meaning 0.16\%${\pm}$0.03\% (52 / 32453) of the main-belt asteroids are A-type. There are 415,448 known objects according to the Minor Planet Center (MPC) within the same semi-major axis and H-magnitude boundaries (as of January 7, 2018) and an A-type abundance of 0.16\%${\pm}$0.03\% results in an estimated 666$^{+134}_{-115}$ A-types with H magnitude $<$ 17 ($\sim$2 km for A-types).  This calculated value is a lower boundary because asteroid discoveries are incomplete at the smaller size ranges particularly at larger distances. We do not attempt in this work to correct for discovery incompleteness in the MPC database. For simplicity, we refer to the total number of A-types as $\sim$600 throughout the rest of the paper.

We calculate the frequency of A-types broken down by region (Inner, Middle, Outer etc), semi-major axis bin, and size (H magnitude). The results are provided in Table~\ref{tab:regions}. We find that half of the $\sim$600 A-types in the main belt with H $<$ 17 are in the smallest size bin 16 $\leq$ H $<$ 17. A-types are roughly evenly distributed throughout the 3 main regions of the main belt (inner, middle, and outer). There is no statistically significant difference between the 3 regions: the fraction of A-types is $0.22^{+0.07}_{-0.06}$\% of the inner belt, to $0.14^{+0.05}_{-0.04}$\% of the middle belt, to $0.11^{+0.06}_{-0.04}$\% of the outer belt. The Hungaria and Cybele regions do not have many candidates, thus those results are more uncertain.

In Table~\ref{tab:obs} we list the albedo for any asteroid when available in the literature. The average for A-types in our sample is 0.28$\pm0.09$ and for comparison the average for S-types in our sample is 0.34$\pm0.11$. The average size of the bodies in these two sets is around 5km, and both sets include 16 objects. The albedos of A-types are statistically indistinguishable from S-types in our dataset.

   \begin{figure}
  \centering
  \includegraphics[width=\textwidth]{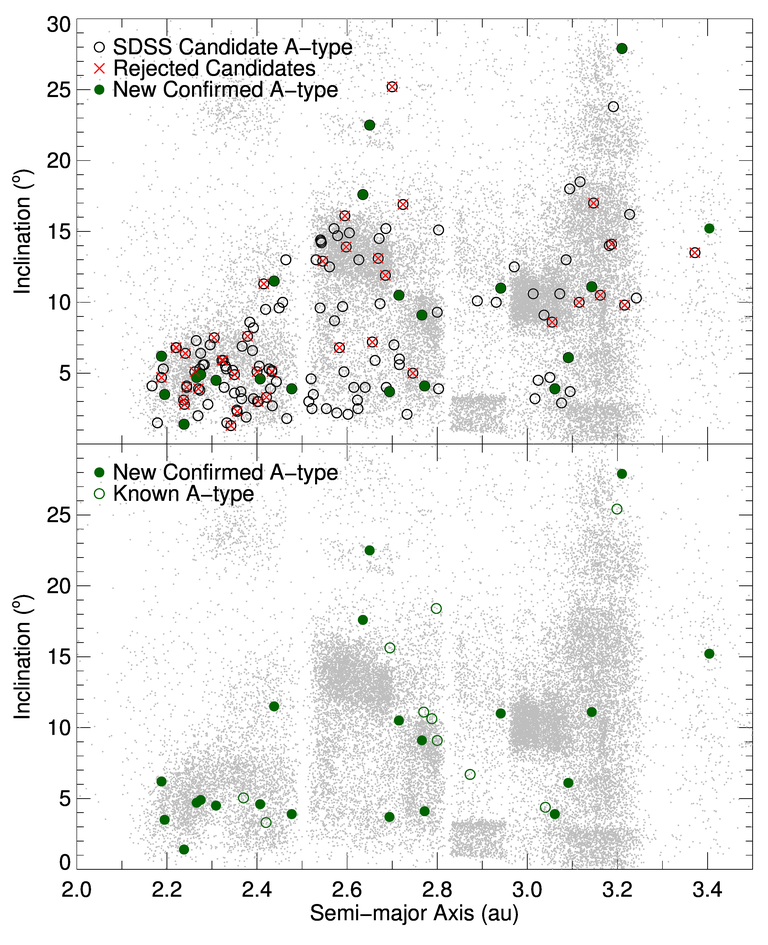}
  \caption[]{%
    	Orbital distribution of A-types in the main belt.  The top panel includes all candidates and survey results, the bottom panel shows only true A-type (olivine-dominated) bodies to highlight their locations. A sample of MBAs are plotted in gray to illustrate the orbital structure of the belt.  SDSS candidates are plotted as black open circles. Green solid circles indicate A-types that have been confirmed with near-infrared data. Red X marks objects with follow up observations that do not classify as A-types. Previously known A-types are plotted as open green circles. It is apparent that A-types exist at all regions of the belt even at high inclinations. We note there is even an A-type past 3.3 AU in the Cybele region where even S-complex asteroids are rare (Sec.~\ref{sec: cybele}). This object, (11616) 1996 BQ2, was discovered in this work as part of the candidates that were outside of the boundaries used for the statistical work here. } 
  \label{fig: orbits}
\end{figure}

\begin{table*}[t]
\begin{minipage}[t]{\textwidth}
\caption{Previously Known A-type Asteroids}
\label{tab:KnownA}
\renewcommand{\footnoterule}{}
\begin{center}
\begin{tabular}{ccccccccccccc}
  \hline
Number\footnote{We only include A-types confirmed with near-infarared data to be olivine-dominated.}	&	Name	&	Orbit\footnote{MB=Main Belt, MC=Mars Crosser, MT=Mars Trojan, Hun=Hungaria} &	H	&	a	&	Ecc	&	Incl	&	Class	&	Class 	&	Albedo	&	Albedo	&	Diam&	Est. Mass\footnote{Mass is estimated as $\frac{4}{3}\pi r^3\rho$. Where the radius is calculated from the H magnitude and the albedo from a survey where marked or 0.2 if unknown (0.2 is the average albedo of A-types in \citet{DeMeo2013} based on survey data), and the density $\rho$ is taken as 2 g/cm$^3$. This density is lower than that calculated for A-types in \citet{Carry2012}, but in the typical range for asteroids, thus keeping our mass estimates conservative. }	\\
	&		&		&	Mag&  (AU)	&		&	(deg)	&		&	Source	&		&	Source	&	(km)	&	(kg)		\\
  \hline
113	&	Amalthea	&	MB	&	8.7	&	2.37	&	0.087	&	5.0	&	Sa	&	SMASS	&	0.2	&		&	53	&	1.57E+17		\\
246	&	Asporina	&	MB	&	8.6	&	2.70	&	0.110	&	15.6	&	A	&	DeMeo	&	0.207	&	WISE	&	55	&	1.76E+17	\\
289	&	Nenetta	&	MB	&	9.5	&	2.87	&	0.204	&	6.7	&	A	&	DeMeo	&	0.291	&	AKARI	&	31	&	3.08E+16		\\
354	&	Eleonora	&	MB	&	6.4	&	2.80	&	0.114	&	18.4	&	A	&	DeMeo	&	0.173	&	WISE	&	165	&	4.66E+18		\\
446	&	Aeternitas	&	MB	&	8.9	&	2.79	&	0.127	&	10.6	&	A	&	DeMeo	&	0.190	&	WISE	&	51	&	1.35E+17		\\
863	&	Benkoela	&	MB	&	9.0	&	3.20	&	0.030	&	25.4	&	A	&	DeMeo	&	0.112	&	WISE	&	62	&	2.53E+17	\\
984	&	Gretia	&	MB	&	9.0	&	2.80	&	0.198	&	9.0	&	Sa	&	SMASS	&	0.2	&		&	46	&	1.05E+17	\\
1600	&	Vyssotsky	&	Hun	&	12.6	&	1.85	&	0.037	&	21.2	&	A	&	Bus	&	0.547	&	AKARI	&	5	&	1.67E+14	\\
1951	&	Lick	&	MC	&	14.1	&	1.39	&	0.060	&	39.0	&	A	&	SMASS	&	0.2	&		&	4	&	9.53E+13		\\
2501	&	Lohla	&	MB	&	12.1	&	2.42	&	0.196	&	3.3	&	A	&	DeMeo	&	0.2	&		&	11	&	1.55E+15	\\
3819	&	Robinson	&	MB	&	11.9	&	2.77	&	0.137	&	11.1	&	Sa	&	SMASS	&	0.2	&		&	12	&	1.99E+15		\\
4490	&	Bamberry	&	Hun	&	13.1	&	1.93	&	0.09	&	26.1	&	A	&	S3OS2	&	0.2	&		&	7	&	3.79E+14		\\
5261	&	Eureka	&	MT	&	16.1	&	1.52	&	0.06	&	20.2	&	A	&	SMASS	&	0.2	&		&	2	&	6.01E+12		\\
7468	&	Anfimov	&	MB	&	12.2	&	3.04	&	0.124	&	4.4	&	A	&	SMASS/Bus	&	0.2	&		&	11	&	1.32E+15		\\
\hline
\end{tabular}
\end{center}
\end{minipage}
\end{table*}

\begin{table*}[t]
\begin{minipage}[t]{\textwidth}
\caption{Calculation of total A-types in the Main Belt}
\label{tab:regions}
\renewcommand{\footnoterule}{}
\begin{center}
\begin{tabular}{lllllllllll} \hline
Boundary	&	
SDSS A-type\footnote{The number of SDSS A-type candidates in each defined bin.}	&	
SDSS A-type \footnote{The number of expected A-types in each bin based on the 33 \% confirmation rate.}	&	
SDSS\footnote{The total number of asteroids (regardless of class) in each bin in our SDSS sample defined from \citet{DeMeo2013}.}	&	
SDSS A-type\footnote{The fraction of each bin that is A-type calculated as N\_expected\_A-type / N\_objects.}	&	
MPC	\footnote{The number of known asteroids in each bin according the to the Minor Planet Center MPCORB.dat dated January 7, 2018.}&	
MPC A-type\footnote{The number of total expected A-types in that bin calculated as MPC\_N\_objects / SDSS\_fraction\_A-type.}	&	
Poisson\footnote{Lower bound on the number of A-types based on Poisson statistics.}	&	
Poisson	&	
Error	\footnote{Error calculated as MPC\_N\_A-type - Poisson\_Lower.}&	
Error	\\
	&	Candidates	&	N Expected	&	N objects	&	Fraction	&	N objects	&	N Expected	&	Lower	&	Upper	&	Lower	&	Upper	\\
\hline H mag\footnote {The first bin spans 12 $\leq$ H $<$ 13 and similar for the other bins. Semi-major axis is bounded in these bins by 1.8 $\leq$ a $<$ 3.7. We do not include larger objects H$<$12 because most asteroids this bright saturated the SDSS detector during observations, meaning the SDSS sample was biased. Known large A-type asteroids are listed in Table~\ref{tab:KnownA}.} \\	
12	&	9	&	3.0	&	1478	&	0.00203	&	4264	&	8.65	&	3.17	&	19.33	&	5.48	&	10.67	\\
13	&	25	&	8.3	&	4568	&	0.00182	&	15180	&	27.69	&	16.27	&	45	&	11.42	&	17.31	\\
14	&	47	&	15.7	&	9377	&	0.00167	&	48111	&	80.38	&	55.76	&	113.04	&	24.62	&	32.66	\\
15	&	48	&	16.0	&	10815	&	0.00148	&	124862	&	184.72	&	128.15	&	259.77	&	56.57	&	75.04	\\
16	&	23	&	7.7	&	5737	&	0.00134	&	221785	&	296.38	&	174.13	&	481.62	&	122.26	&	185.24	\\
\hline Region\footnote{Regions are defined as Hungaria (1.8 $\leq$ a $<$ 2.0 AU and i $>$ 15), Inner Belt (2.1 $\leq$ a $<$ 2.5 AU), Middle Belt  2.5 $\leq$ a $<$ 2.82 AU, Outer Belt (2.82 $\leq$ a $<$ 3.3 AU), Cybele (3.3 $\leq$ a $<$ 3.7 AU) with 11 $\leq$ H\_magnitude $<$ 17 for all regions.} \\
Hun	&	3	&	1.0	&	379	&	0.00264	&	2451	&	6.47	&	0.65	&	25.22	&	5.82	&	18.75	\\
Inn	&	70	&	23.3	&	10740	&	0.00217	&	83549	&	181.52	&	134.95	&	240.71	&	46.56	&	59.19	\\
Mid	&	51	&	17.0	&	12072	&	0.00141	&	148869	&	209.64	&	147.98	&	291.03	&	61.66	&	81.39	\\
Out	&	29	&	9.7	&	9118	&	0.00106	&	177897	&	188.6	&	116.93	&	290.45	&	71.67	&	101.84	\\
Cyb	&	1	&	0.3	&	139	&	0.00240	&	2433	&	5.83	&	0.58	&	22.75	&	5.25	&	16.92	\\
\hline Semi-major\footnote{H magnitude is bounded in these bins by 11 $\leq$ H\_magnitude $<$ 17.} \\
Axis (AU) \\
2.2	&	21	&	7.0	&	3206	&	0.00218	&	21213	&	46.32	&	25.8	&	78.08	&	20.51	&	31.76	\\
2.3	&	23	&	7.7	&	4445	&	0.00172	&	35702	&	61.58	&	36.18	&	100.06	&	25.4	&	38.49	\\
2.4	&	20	&	6.7	&	2458	&	0.00271	&	22640	&	61.4	&	34.21	&	103.51	&	27.19	&	42.11	\\
2.5	&	21	&	7.0	&	3945	&	0.00177	&	39275	&	69.69	&	38.83	&	117.48	&	30.86	&	47.79	\\
2.6	&	17	&	5.7	&	4435	&	0.00128	&	54822	&	70.05	&	36.19	&	122.58	&	33.86	&	52.54	\\
2.7	&	10	&	3.3	&	3387	&	0.00098	&	49376	&	48.59	&	17.82	&	108.53	&	30.78	&	59.93	\\
2.8	&	4	&	1.3	&	1050	&	0.00127	&	18158	&	23.06	&	2.31	&	89.93	&	20.75	&	66.87	\\
2.9	&	3	&	1.0	&	1617	&	0.00062	&	24380	&	15.08	&	1.51	&	58.8	&	13.57	&	43.72	\\
3.0	&	13	&	4.3	&	2912	&	0.00149	&	56499	&	84.08	&	35.73	&	168.15	&	48.34	&	84.08	\\
3.1	&	8	&	2.7	&	3110	&	0.00086	&	70667	&	60.59	&	22.22	&	135.33	&	38.38	&	74.73	\\
3.2	&	4	&	1.3	&	734	&	0.00182	&	13589	&	24.68	&	2.47	&	96.27	&	22.22	&	71.59	\\
3.3	&	1	&	0.3	&	68	&	0.00490	&	1014	&	4.97	&	0.5	&	19.39	&	4.47	&	14.41	\\

\hline

\end{tabular}
\end{center}
\end{minipage}
\end{table*}

\subsection{An A-type among Cybeles} \label{sec: cybele}
We discovered an A-type (11616) 1996 BQ2 with a semi-major axis of 3.4 AU and inclination of 15 degrees, a location that places it within the Cybele region (3.3 $\leq$ a $<$ 3.7 AU), even further than the outer main belt 2.82 $\leq$ a $<$ 3.3 AU). This A-type was discovered outside of the formal statistical survey presented here, because its slope was too low (gri-slope 1.98, z-i value -0.15) to be within the formally defined A-type boundaries (see Sec~\ref{sec: targetselection}). The spectrum is shown in the right panel of Fig.~\ref{fig: famspec}. SDSS data for asteroid (11616) was classified as an S-type by \citet{Carvano2010}, \citet{Gil-Hutton2010}, and \citet{DeMeo2013}. \citet{Popescu2016} indicated that it was an A-type candidate based on near-infrared colors (two to four broad band filters per observation) from the VISTA Hemispheric Survey (VISTA-HMS) catalog that included 40,000 asteroids, 146 of which were Cybele.

The Cybele region is dominated by taxonomic classes considered to have formed in colder environments, further from the Sun, such as C-, P-, and D-type asteroids. S-types were thought to have formed in warmer environments closer to the Sun and while they are not compositionally equivalent to A-types, they can be used as a basis for comparison for the distribution of these types of bodies. S-types make up 21, 8, and 5\% of the inner, middle and outer belt by mass, whereas only 1\% of the Cybeles are S-type \citep{DeMeo2013,DeMeo2014}. Finding an A-type in a region where S-types are rare is an interesting result that can place constraints on how and when A-types were delivered (or formed) in the main belt. 

\subsection{Family Results} \label{sec: fam}
We observed 33 A- and V-type candidates within or near asteroid families. Twelve candidates are unique to this study and are not part of the general A-type distribution survey. The confirmed A- and V-type spectra are plotted in Fig.~\ref{fig: famspec}. The full set of spectra are plotted in Supplementary Figure 5. A summary of the results for each family is presented in Table~\ref{tab:famsum}. Note we did not search for V-types in any families in the inner belt. The presence of the Vesta family and the large number of V-types associated with that family would make it difficult to unambiguously link a V-type in the inner belt with any other family. 

We find one V-type in the Eunomia family, asteroid (66905) (see Fig~\ref{fig: famspec}). We find one A-type asteroid, (17818), in the Gefion family, and one, (92516), near the Vesta family (although it is formally outside the family boundaries defined by \citet{Nesvorny2010}). We find three A-types dynamically linked to the Flora family, (16520), (139045), and (34969). Asteroid (16520) is also dynamically within the Baptistina family. In Fig~\ref{fig: famorb} we show a ``butterfly'' plot of the Flora family with semi-major axis and H magnitude showing how the family spreads in distance with smaller size due to the Yarkovsky effect \citep{Burns1979,Bottke2006Yorp}. Two of the three A-types are relatively small, making it more challenging to rule out that they could be interloper, background bodies. Family membership is more substantively addressed in Section~\ref{sec: dynamics}.

We perform a statistical analysis of the differentiated fragments confirmed within families to determine if the frequency within any family is significantly different than for the general main-belt A-type population determined from Sec~\ref{sec: aresult}. Results are provided in Table~\ref{tab:famsum}. Binomial statistics are calculated with n (number of trials), v (number of successes), and p (probability of success) where n is the family total number of objects, v is the estimated number of A- or V-types in the family rounded to a whole number, and p is 0.0016, calculated as the confirmation rate of A-types in the general main-belt population as 52 / 32453 (see Sec~\ref{sec: aresult}). The cumulative probability is given as the Probability P that the number of A- or V-types is less than the observed number, P(X $<$ v).

Even though the fraction of A-types in the Flora and Gefion family is higher than the general main-belt A-type population (0.41\% and 0.57\% respectively compared to 0.16\%), we find there is an 88.2\% and 84.6\% chance that there should be fewer A-types in the Flora and Gefion families respectively than we find for the general A-type population. Neither of these results raise to the level of statistical significance, thus we find no compelling evidence for differentiation within families.

   \begin{figure}
  \centering
  \includegraphics[width=\textwidth]{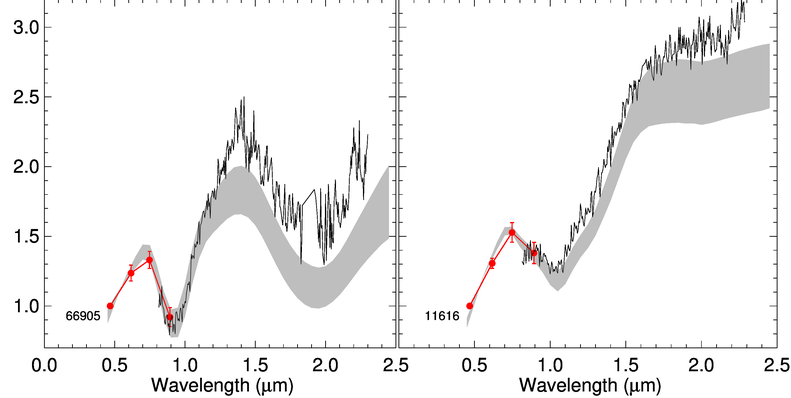}
  \caption[]{%
    	Left: Spectrum of an observed V-type with an orbit within the Eunomia family (Sec.~\ref{sec: fam}). The spectra of confirmed A-types within families are plotted within Fig.~\ref{fig: specA}. The spectra of all observed candidates, including false positives are provided in Supplementary Figure 5. Right:  Asteroid (11616) is an A-type discovered within the Cybele region of the asteroid belt (Sec.~\ref{sec: cybele}). This asteroid was part of the set of observations of potential A-types that were not within the formal A-type candidate boundaries. } 
  \label{fig: famspec}
\end{figure}

\begin{table*}[t]
\begin{minipage}[t]{\textwidth}
\caption{Summary for the Survey of Family Candidates}
\label{tab:famsum}
\renewcommand{\footnoterule}{}
\begin{center}
\begin{tabular}{lllllllll}
  \hline
  
Family	& Family\footnote{Number of objects in the family as identified by \citet{Nesvorny2010} that have classifications in \citet{DeMeo2013}.} & 
Num \footnote{Number of A- or V-type candidates identified in the SDSS sample.} & 
Num	\footnote{Number of A- or V-type candidates observed.}&	
Confirmed \footnote{Eunomia had one confirmed V-type all other confirmations were A-type.} 	&	
Est Num\footnote{Calculated as Family\_Total * (N\_Confirmed / N\_Observed).}&	
A or V as\footnote{Calculated as Est\_Num\_in\_fam / Family\_Total.}	&	
Binomial\footnote{Binomial statistics are calculated with n (number of trials), v (number of successes), and p (probability of success) where n is Family\_tot, v is Est\_Num\_in\_Fam rounded to a whole number, and p is 0.0016, calculated as the confirmation rate of A-types in the general main-belt population as 52/ 32453 (see text in Sec~\ref{sec: aresult}). The cumulative probability is given as the Probability P that the number of A- or V-types is less than the observed number, P(X $<$ v).}	&	
Binomial\footnote{The exact binomial probability P is where P(X = v).}	\\
	& Total &Cands	&Obs.	&	A- or V-type	&in Fam	&	\% of Fam	 &	Cumulative (\%) &	Exact (\%)	\\

\hline
Eos	&	1010	&	3	&	1	&	0	&	0.00	&	0	&		&		\\
Eunomia	&	1124	&	18	&	12	&	1	&	1.50	&	0.1335	&	16.7	&	26.8	\\
Flora	&	1166	&	19	&	12	&	3	&	4.75	&	0.4074	&	88.2	&	2.9	\\
Gefion	&	437	&	5	&	2	&	1	&	2.50	&	0.5721	&	84.6	&	2.8	\\
NysaPolana	&	1080	&	7	&	3	&	0	&	0.00	&	0	&		&		\\
Vesta	&	1360	&	5	&	3	&	1	&	1.67	&	0.1225	&	11.5	&	26.9	\\
\hline
\end{tabular}
\end{center}
\end{minipage}
\end{table*}

   \begin{figure}
  \centering
  \includegraphics[width=0.8\textwidth]{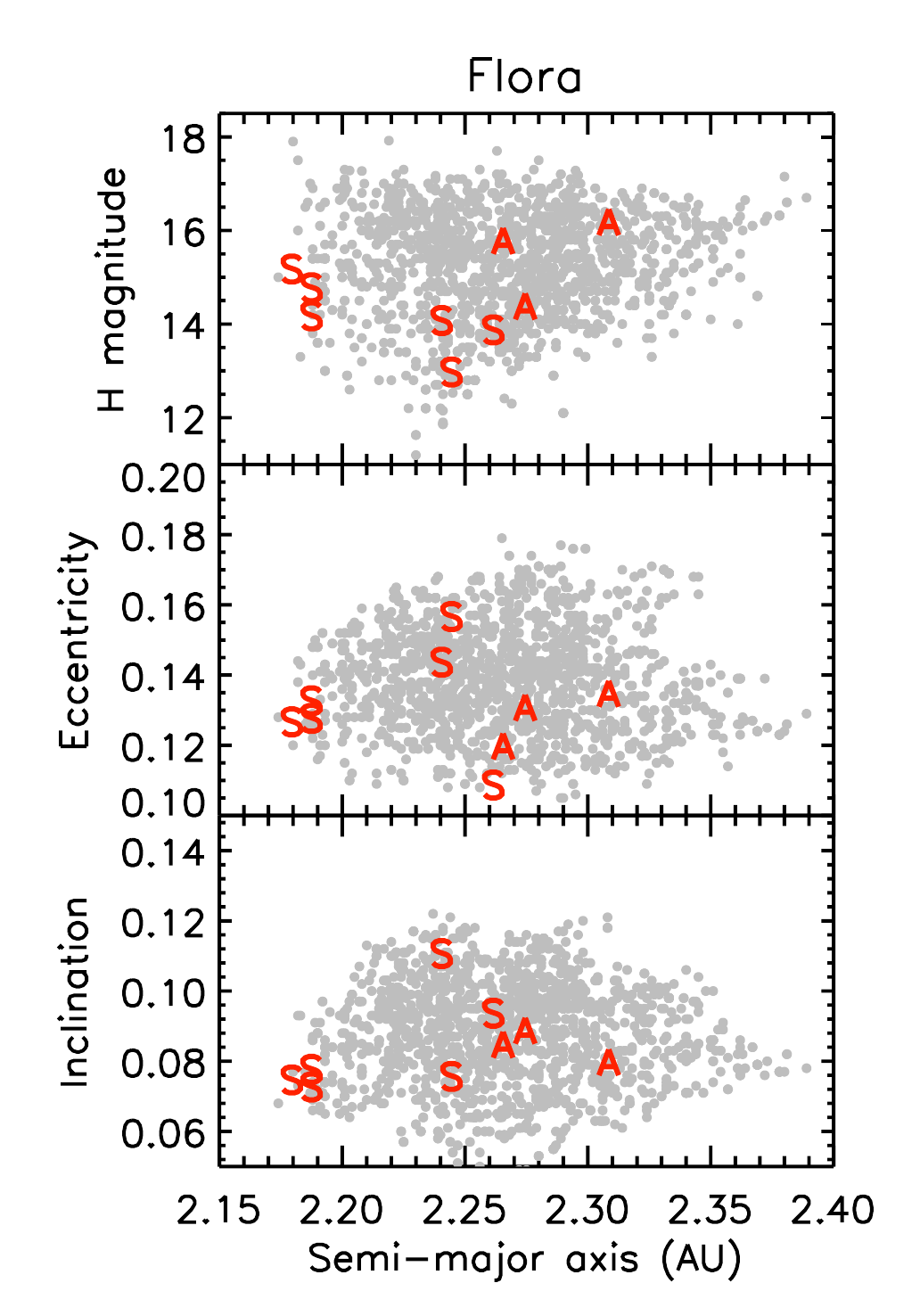}
  \caption[]{%
    	Plot of the Flora family as semi-major axis versus H-magnitude (top), proper eccentricity (middle), and proper inclination (bot).	Gray dots are Flora family members defined in \citet{Nesvorny2010}, red letters are the classifications of bodies observed in this work. } 
  \label{fig: famorb}
\end{figure}

\subsection{Family Membership: A Closer Look} \label{sec: dynamics}

A challenge in identifying differentiation within families is the ever-present potential for interlopers caused by the limitations of dynamical methods for distinguishing between family members and background objects, especially for large and old families (which are typically highly dispersed in orbital element space) in densely populated regions of the asteroid belt \citep{migliorini1995_interlopers}.  The Flora family is a good example of such a challenging case, as it is located in a crowded region of the inner asteroid belt near several other families, including the Baptistina, Vesta, Massalia, and Nysa-Polana families \citep{dykhuis2014_flora}, is considered unusually dispersed in eccentricity and inclination relative to other asteroid families \citep{nesvorny2002_flora}, and is affected by a number of nearby or crossing dynamical resonances \citep{vokrouhlicky2017_flora}.  As such, membership lists for the Flora family constructed using the widely-employed Hierarchical Clustering Method (HCM) for family identification \citep{zappala1990_hcm,zappala1994_hcm} are suspected to have a significant fraction (perhaps as high as 50\%) of interlopers from neighboring families and the background population \citep[e.g.,][]{migliorini1995_interlopers,dykhuis2014_flora,oszkiewicz2015_flora}.

Consideration of compositional information, such as colors or albedos, is often included in family classification efforts to help remove interlopers from lists of family members \citep[e.g.,][]{novakovic2011_highifamilies,masiero2013_astfams_neowise}, under the assumption that family members should be relatively compositionally homogeneous having originated from the same parent body, assuming that the parent body itself was compositionally homogeneous \citep[e.g.,][]{ivezic2002_astfamilies,cellino2002_astfamilyspectroscopy_ast3}.  By removing compositional outliers, however, this approach then necessarily limits our ability to search for taxonomic variability among objects considered to be ``true'' members of a particular family.

To assess the likelihood independent of compositional considerations that the A- and V-type objects that we find within or near asteroid families could be interlopers, we perform a simple analysis based on dynamical methods and considerations alone.  Focusing on the Flora family, where we find three A-type asteroids among the family members we observed, we perform dynamical integrations of all 12 Flora family members we studied and an additional twenty of the lowest numbered asteroids in the family to characterize their dynamical behavior over time.  We generate four dynamical clones per object, which combined with each original object, gives a total of five test particles per object, where the dynamical clones are Gaussian-distributed in orbital element space (characterized by $\sigma$ values of 1$\times$10$^{-6}$~AU for semimajor axes, 1$\times$10$^{-5}$ for eccentricities, and 1$\times$10$^{-4}$${\degr}$ for inclinations) and centered on each object's osculating orbital elements as of 2017 January 1.  The sigma values for the clones are very conservative. For example asteroid (139045) has 1 sigma errors on a, e, i of 1$\times$10$^{-8}$, 4$\times$10$^{-8}$, and 6$\times$10$^{-6}$, respectively, orders of magnitude smaller than assumed here. We then perform forward integrations for 100 Myr using the Bulirsch-St{\" o}er integrator in the Mercury numerical integration software package \citep{chambers1999_mercury}. We include the gravitational effects of all eight major planets and treat all test particles as massless. Non-gravitational forces are not considered in this analysis.

We find that all test particles associated with observed Flora family members remained stable against ejection from the solar system (defined as occurring when the semimajor axis, $a$, of an object exceeds 100~AU) over the entire integration period.  One dynamical clone of one of the A-type asteroids found in the family (16520) underwent an excursion in semimajor axis of $\Delta a\sim0.025$~AU over the course of the integration, but all other test particles associated with A-type asteroids remained extremely stable, undergoing maximum excursions of $\Delta a<0.005$~AU.  The vast majority of the other test particles associated with observed Flora family members exhibited similar stable behavior, although three dynamical clones of non-A-type asteroids underwent relatively large excursions in semimajor axis ($\Delta a>0.05$~AU) over the course of the integrations. Meanwhile, all test particles with the exact osculating elements of the 20 low-numbered Flora family members that we did not observe also remained stable over the duration of our integrations, although nine of the dynamical clones of these objects were ejected during our integrations.  We construct contour plots showing the relative density of intermediate positions (in time steps of 10,000 years) in orbital element space over the duration of our integrations occupied by all test particles (i.e., original objects and dynamical clones) associated with the three A-type asteroids found in the Flora family, as well as (8) Flora itself for reference.  We find the dynamical evolution of all three A-type asteroids to be qualitatively similar to that of Flora as well as most of the other Flora family members for which we also performed dynamical integrations.

\begin{figure}
  \centering
  \includegraphics[width=0.8\textwidth]{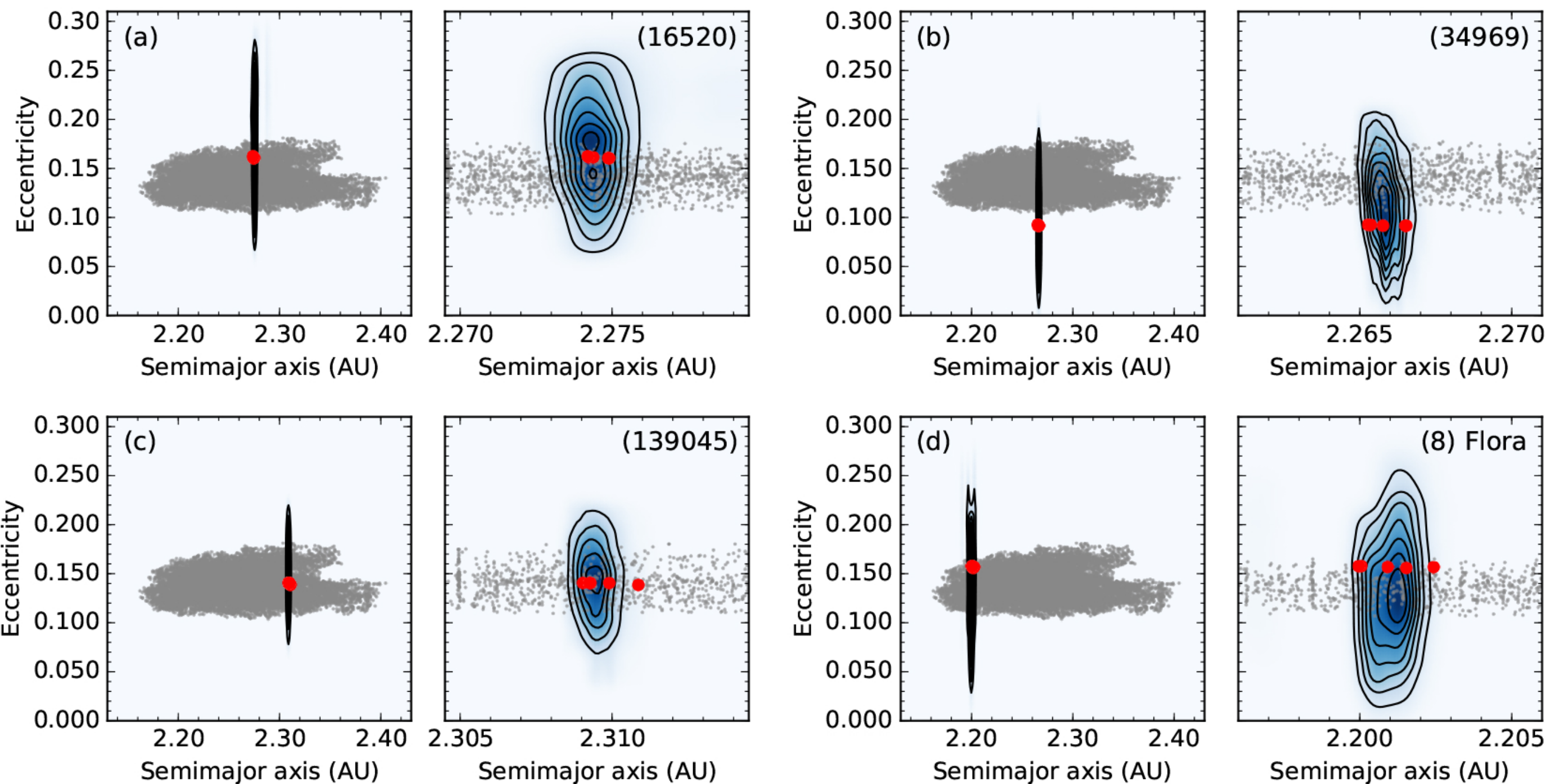}
  \caption[]{Contour plots (black lines) showing the relative density of intermediate positions (in time steps of 10,000 years) in orbital element space occupied by all test particles (i.e., original objects and dynamical clones) associated with A-types (a) 16520, (b) 34969, (c) 139045, and (d) (8) Flora (the largest member and namesake of the Flora family) during the 100~Myr integrations described in the text. The left panel of each set of plots shows wide views (in terms of semimajor axis range) of each object's dynamical evolution with respect to the entire Flora family (synthetic proper elements for all family members identified by \citet{nesvorny2015_pdsastfam} plotted with small gray dots for reference) and the right panel shows more detailed views of the regions of orbital element space occupied by the test particles for each object during the integrations.  Darker shaded blue areas indicate higher densities of intermediate orbital element positions.
} 
  \label{fig:integrations}
\end{figure}

\begin{figure}
  \centering
  \includegraphics[width=0.8\textwidth]{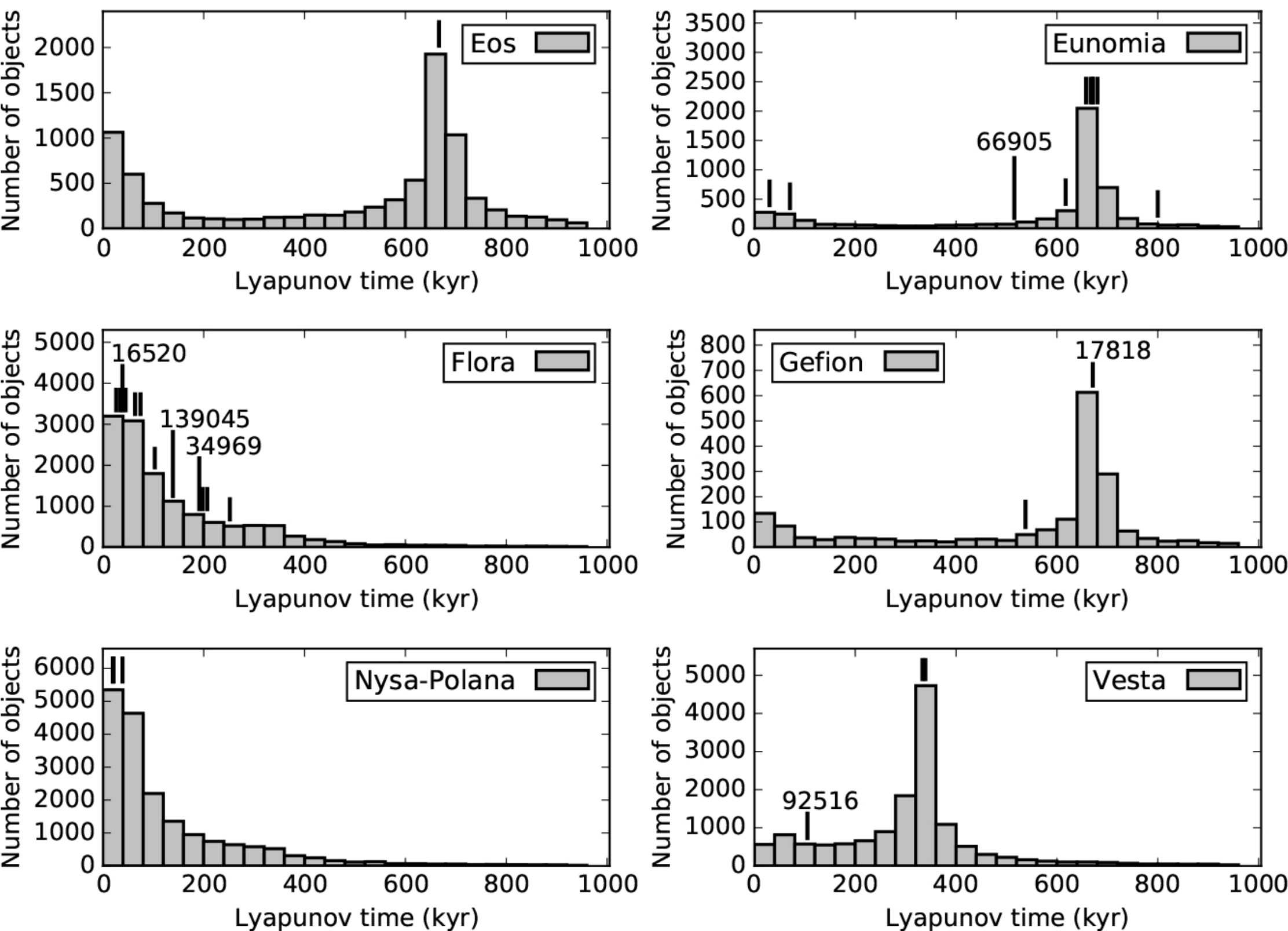}
  \caption[]{Histograms of Lyapunov times in kyr from the AstDyS website for asteroids identified as members of the Eos, Eunomia, Flora, Gefion, Nysa-Polana, and Vesta families, as labeled, by \citet{nesvorny2015_pdsastfam}.  Lyapunov times for A-type asteroids found in the Flora, Gefion, and Vesta families and the V-type asteroid found in the Eunomia family in this work are marked with vertical line segments and numerical labels.  Other vertical line segments mark Lyapunov times for other family-associated objects observed in this work.
} 
  \label{fig:hist_lyapunov_times}
\end{figure}

Long-term dynamical stability does not necessarily guarantee that an object is a true member of an asteroid family, as there is no a priori condition that requires members of asteroid families to be dynamically stable.  However, if we had found short dynamical lifetimes for the A-type asteroids we find in the Flora family, it would have suggested that these objects were likely to have been recently delivered to their current locations, therefore increasing the likelihood that they could be interlopers, especially if other members of the family are significantly more dynamically stable.  That said, given the numerous dynamical resonances that cross the region, even true members of the Flora family may currently have short dynamical lifetimes due to the Yarkovsky effect nudging them towards more unstable regions over time.  Indeed, nearly 50\% of Flora family members identified by \citet{nesvorny2015_pdsastfam}, including one of the A-type asteroids we found in the family (16520), have Lyapunov times of $t_{ly}<100$~kyr (typically the threshold under which an object is considered dynamically unstable), according to a synthetic proper element catalog retrieved from the AstDyS website\footnote{{\tt http://hamilton.dm.unipi.it/astdys/}} on 2018 January 1 (Figure~\ref{fig:hist_lyapunov_times}).  Combined these results with those of our numerical integrations, we conclude that we find no indications of anomalous dynamical behavior by any of the A-type asteroids we find in the Flora family, and thus no compelling dynamical evidence that they are likely to be interlopers.

Given that we only find one compositionally anomalous (i.e., A- or V-type) asteroid per family in the Eunomia, Gefion, and Vesta families, we did not perform full dynamical integrations for objects in those families.  We note, though, that similar to the A-type asteroids we find in the Flora family, none of these asteroids (and in fact, none of the family-associated asteroids that we observed at all) have $t_{ly}$ values that are anomalous for the families with which they are associated (Figure~\ref{fig:hist_lyapunov_times}).  As such, while these results should not be regarded as definitively establishing that these objects are true members of their respective families, we nonetheless conclude for now that we find no anomalous dynamical behavior with respect to other family members that might suggest that these objects are likely to be interlopers. As such, we conclude for now that we find no anomalous dynamical behavior with respect to other family members that might suggest that these objects are likely to be interlopers.  We note though that Yarkovsky drift could cause family members to eventually evolve onto orbits similar to those of a nearby A-type background object, or vice versa, thus making the family members and the A-type object effectively indistinguishable from each other using dynamical criteria, even though the A-type object is an interloper.  Therefore, as before, while these results are suggestive, they should not be interpreted as definitively establishing that these objects are true members of their respective families.

More sophisticated methods have been employed in attempts to distinguish true members of a family from interlopers, such as the modeling of the dynamical evolution (including the Yarkovsky effect) of suspected interlopers from their assumed origin points \citep[e.g.,][]{carruba2005_vesta}, or the use of a family's ``core'' in orbital elements (where the fraction of interlopers is assumed to be relatively low) to establish the range of reflectance properties of the family before expanding consideration to the rest of the family \citep[e.g.,][]{dykhuis2014_flora}.  If and when individual interlopers can be reliably identified, an automated method has also been developed to remove additional associated interlopers that were linked to the family through the use of HCM via ``chaining'' through the initially identified interloper \citep{radovic2017_interlopers}.  Such detailed analyses of each of the families we studied here is beyond the scope of this work, particularly given the low incidence of A-type asteroids found in the families we studied, and thus the minimal evidence of differentiation in those families.  Nonetheless, in the event that stronger evidence of differentiation in these or other families is found in the future, these methods may be worth considering to ensure that any compositional diversity found within a family is truly due to differentiation of the parent body and not subsequent contamination by interlopers.

\subsection{A note on observational completeness} \label{sec: completeness}

Correcting for observational incompleteness is beyond the scope of this work. However, here we explain the completeness factors to consider.  These questions include 1) How complete is the SDSS at a given H magnitude? 2) How complete is the MPC at the same magnitude? 3) How does one account for the fact that a 2 km body has a different average H magnitude depending on its taxonomic class? In \citet{DeMeo2013} and \citet{DeMeo2014} an upper H magnitude limit of 15.5 was chosen because the SDSS survey was sensitive to even the darkest asteroid class at that magnitude.

In this work we are pushing down to 2 km (H$\sim$15.9 for A-types and H$\sim$17.2 for D-types) to increase our candidate pool of A-types, the SDSS survey cannot measure darkest asteroids at further distances. This means that SDSS, the MPC catalog of known asteroids (and thus our results) are more incomplete at the smallest size ranges, but are differently incomplete for each class.

For example, the SDSS is biased towards observing the high-albedo A-types over the darker classes. Therefore, we are overestimating the fraction of A-types relative to other classes in SDSS, excluding the inner main belt which is sensitive down to H=17.2, thus it includes all 2 km bodies regardless of class. On the other hand, because the MPC is incomplete at smaller sizes we are underestimating the total number of bodies. This comes into effect when we apply the fractions from SDSS to the MPC numbers to determine the total number of A-types at each distance and size range. For example, in 2013 we estimated that in the H = 15-16 magnitude range, we were 100\%, 85\%, and 60\% complete in the inner, middle, and outer main belt respectively.

\section{Discussion}

\subsection{Discussion on general distribution and mass}
\begin{figure}
  \centering
  \includegraphics[width=0.8\textwidth]{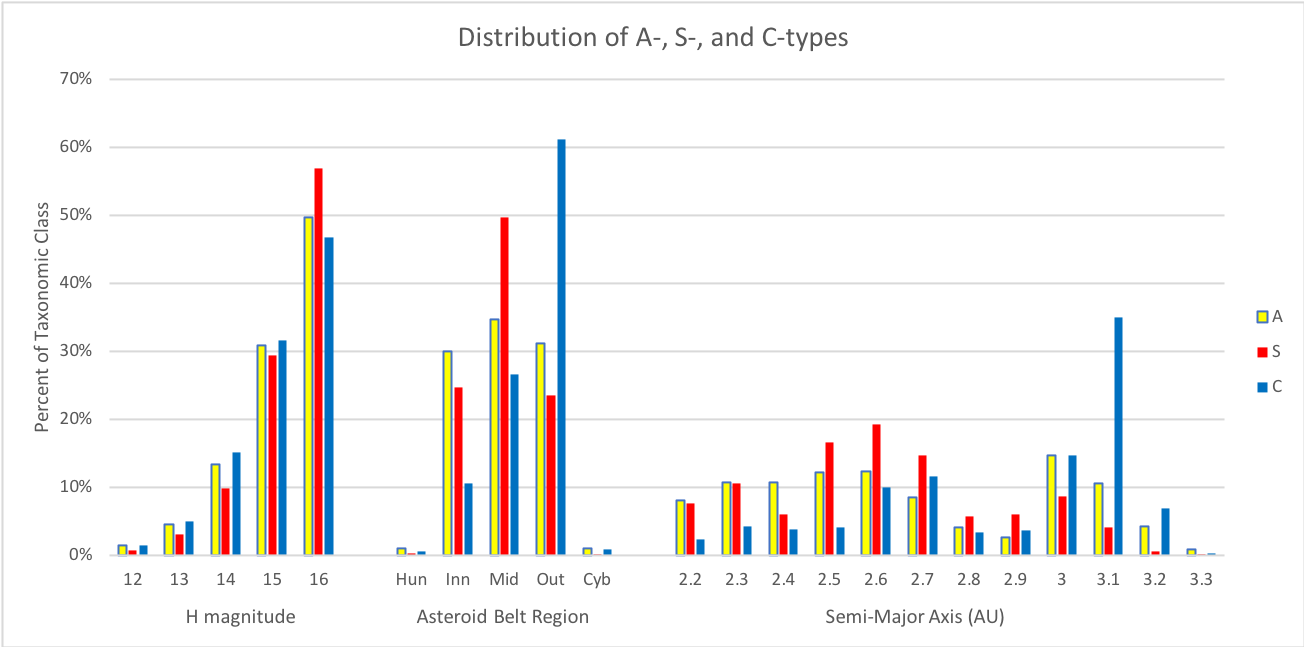}
  \includegraphics[width=0.45\textwidth]{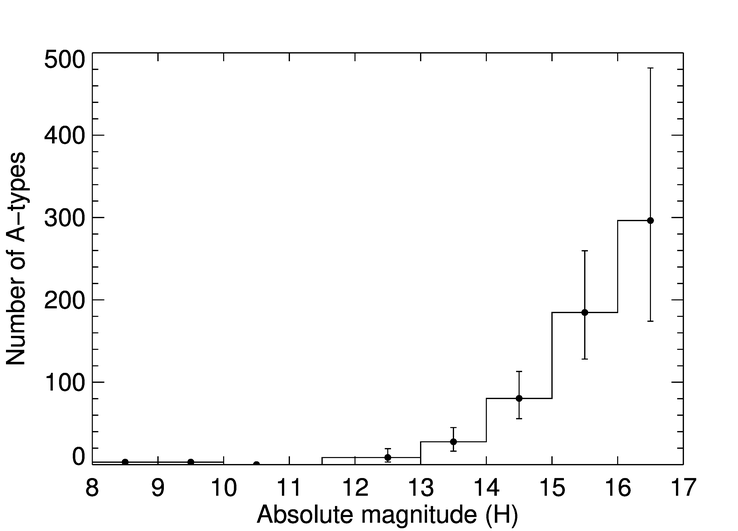}
  \includegraphics[width=0.45\textwidth]{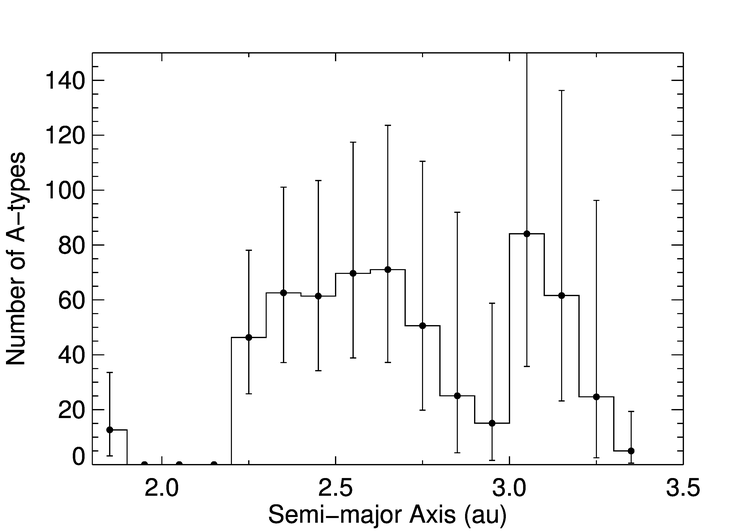}
  \caption[]{%
    	Top: The fractional distribution of the number of A-, S-, and C-types in each bin for H magnitude, region of the belt, and semi-major axis bin. For the main-belt regions, the distribution of A-types is generally consistent with that of S-types, although the A-type distribution is flatter, and is not consistent with the distribution of C-types. Bottom Left: The expected number of A-types in the main belt in each H magnitude bin. Bottom Right: The expected number of A-types greater than $\sim$2 km (H$<$17) as a function of semi-major axis. } 
  \label{fig: histos}
\end{figure}

Overall, we find A-types well distributed throughout the main belt in distance and inclinations suggesting there is no single main-belt common origin. In Fig.~\ref{fig: histos} we show the number of A-types as a function of size and of semi-major axis. Even though A-types are more numerous at smaller sizes, they remain generally the same fraction of the population - as the population increases at smaller sizes so does the number of A-types. We also look at how the A-type sample is distributed across the main belt compared to the S- and C-types. A similar distribution could indicate a similar delivery mechanism. For S and C we use the number of bodies in each bin from the SDSS dataset classified by \citet{DeMeo2013}, and calculate the fraction by dividing by the total number of S and C for each sample (bounded by H magnitude, main-belt region, or semi-major axis) such that the percentage of each type sums to 100\% over the whole region. For the main-belt regions, the distribution of A-types is generally consistent with that of S-types, although the A-type distribution is flatter since S-types peak in the middle belt. The A-type distribution is not consistent with the C-types that make up only a small fraction of the inner belt and increase dramatically, peaking in the outer belt. 

\begin{figure}
  \centering
  \includegraphics[width=\textwidth]{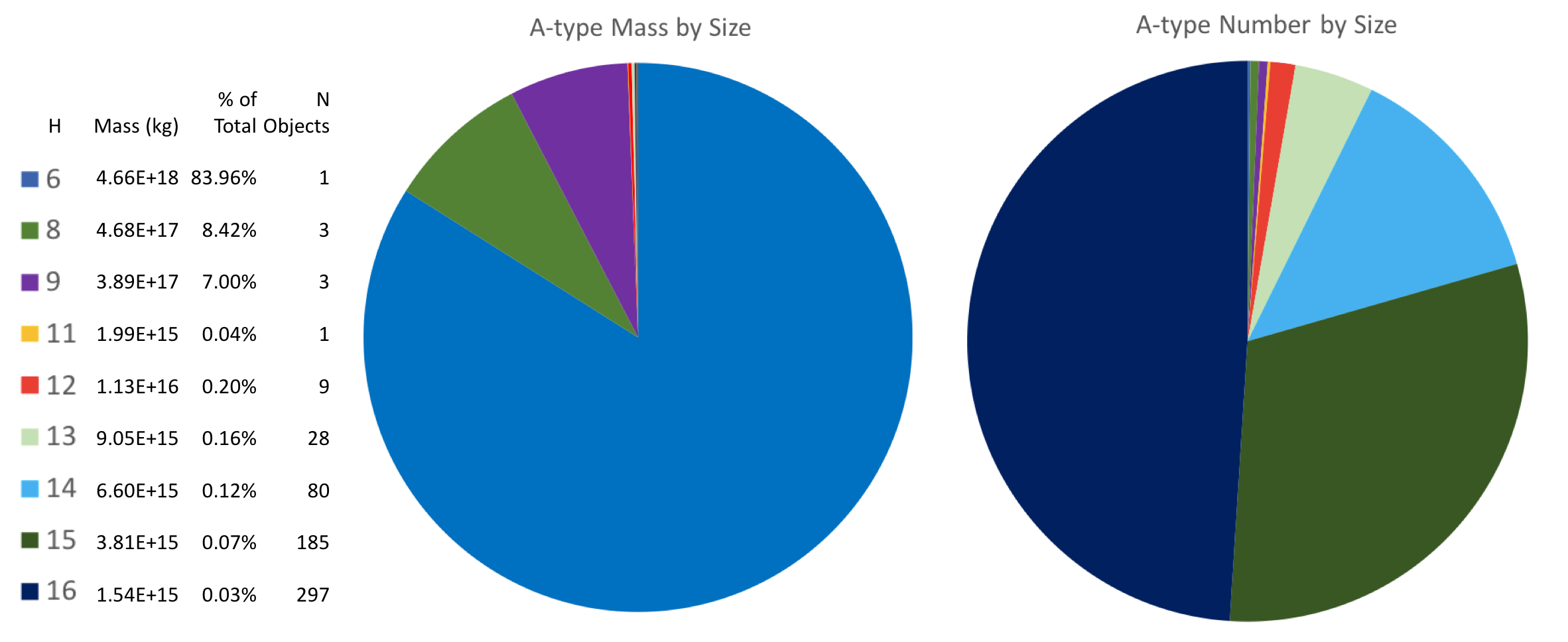}
  \caption[]{%
 The mass and number of A-types in each H magnitude range. The single largest A-type (354) Eleonora accounts for over 80\% of the mass of all A-type asteroids. However, the interior of an asteroid may not be representative of the spectrally measured exterior, particularly for a body as large as Eleonora. While A-types are most numerous at smaller sizes, with about half of them in the smallest size bin (16 $\leq$ H $<$ 17), those smaller bodies comprise an insignificant amount of mass. } 
  \label{fig: mass}
\end{figure}

Most of the mass of A-types ($>$80\%) is contained within the single largest body, (354) Eleonora, that has an absolute (H) magnitude of 6.4 and a mass of roughly 4.6 x 10$^{18}$kg. See \citet{Gaffey2015} for an in-depth spectral analysis of Eleonora. We show the mass of A-types in each size range in Fig.~\ref{fig: mass}. The total mass of A-types we find here is comparable to that calculated in \citet{DeMeo2013} that uses densities from \citet{Carry2012}, and differences can be attributed to the assumptions for density and diameter for the mass calculation and to a better understanding of the false positive rate. For the size range we sample in this survey (12 $\leq$ H $<$ 17 or diameters of approximately 2 $\leq$ D $<$12 km), the total A-type mass in the main belt (according to asteroids discovered through January 2018) is 3 x 10$^{16}$kg, two orders of magnitude less than Eleonora itself. In our sample alone, the mass distribution for each H magnitude bin is 35, 28, 20, 12 and 5\% respectively for H=12, 13, 14, 15, 16.

The focus of this paper is on implications for differentiation among small bodies, however, not all olivine-dominated bodies are differentiated \citep{Burbine2014}. A nebular (primitive) origin is expected for bodies with ferroan olivine compositions (fayalite Fe$_2$SiO$_4$), and a magnesian olivine (forsterite Mg$_2$SiO$_4$) composition indicates differentiation. Using a Modified Gaussian Model (MGM) \citet{Sunshine2007} studies 9 large A-type asteroids measuring the 1-$\mu$m band parameters. She finds 7 of 9 in her study are differentiated and 2 of 9 (289 Nenetta and 246 Asporina) are of nebular origin. This indicates that only 80\% of A-types are differentiated, and any estimate of actual differentiated material should be corrected by this factor. Future work will perform similar models of the spectra from this survey to better constrain the fraction that are differentiated versus nebular. The overall results we find here should remain qualitatively the same and general conclusions still valid even if a fraction are found to be primitive.

\subsection{A-types in the NEO and Mars Crosser Populations}

In a visible-wavelength survey \citep{Perna2018} of near-Earth objects \citet{Popescu2018} detected 8 A-type objects out of a sample of 147. \citet{DeMeo2009taxo} find in their spectral sample that 50\% of visible-wavelength A-types remain A-type with near-infrared data (see Table 3 from that work). Even factoring in a lower success rate the number of A-types in the work by Popescu represents $\sim$2.5\% of their sample, an order of magnitude higher than our finding that 0.16\% of main-belt objects are A-type. Following the SMASS (Small main-belt asteroid Spectral Survey) came the MIT-Hawaii NEO Spectral Survey or MITHNEOS, results of which are published in \citet{Binzel2018}. MITHNEOS compiled 1047 spectrally classified NEOs from their data and the literature, 727 of which have near-infrared spectra. Popescu et al.'s A-type sample are all small NEOs, with H$>$20. Even though the MITHNEOS sample spans a broader H range (primarily between 16 $\leq$ H $<$ 22), the sample size of near-infrared spectra with H $>$ 20 is 162, comparable to that of \citet{Popescu2018}. MITHNEOS observed one of the A-type NEOs in Popescu et al.'s sample (444584) which was classified as an Sq-type based on diagnostic near-infrared data.

Six asteroids in MITHNEOS were classified as A- or Sa-type. Four of those classifications - for asteroids (6053), (275677), (366774), and 1993 TQ2 - were based on visible-wavelength data, two with spectra and two with colors (\citep{Binzel2004,Ye2011,DeMeo2013,Kuroda2014}). Two asteroids, (5131) and 2014 WQ201, with near-infrared spectra are classified as Sa in \citet{Binzel2018}. The quality of the data for 2014 WQ201 was poor and had it been a part of this survey it would not have been classified as confidently olivine-dominated. Our preferred calculation, including (5131) and excluding 2014WQ201 based on data quality results in an olivine-dominated percentage of the NEO population of 0.14\% (1/727) which is a striking match with the main-belt numbers we find here. Even including the second NEO with the poor-quality data results in a value of 0.28\%. Even the most optimistic fraction from the near-infrared MITHNEOS data set looking only at H$>$20 to be consistent with \citep{Popescu2018} is 1/162 or 0.6\%.

Among the Mars Crossers in the MITHNEOS data set (1951) Lick is a well-studied A-type \citep{deleon2004}. One more near-infrared spectrum was classified as A-type, asteroid (367251), however, the quality of this spectral data was very poor. Two additional Mars Crossers were classified as A- or Sa-type base on visible-wavelength data originally published in \citet{Binzel2004}. 

A-types have also been discovered in the Mars Trojan population. Asteroid (5261) Eureka has been known to have a unique spectrum \citep{Rivkin2007}, and even though it has the characteristic red spectrum and wide, deep 1-$\mu$m absorption band of an A-type it is distinct from all other A-types because the 3 minima of the 1$\mu$m band are deeper making the whole feature more bowl-shaped than V-shaped. \citet{Borisov2017} and \citet{Polishook2017} performed near-infrared spectroscopic measurements of a few other Mars Trojans in the same cloud and found that a number of them had spectra similar to that of Eureka, suggesting they are a mini-family and are fragments of a parent body that may have been disrupted in that location. \citet{Polishook2017} put forth the theory that these fragments could have originated from Mars itself, following a large impact (such as the one that formed the Borealis basin) that excavated olivine-rich material from the Martian mantle. Their dynamical model also showed that A-types found in the Hungaria family might also have resulted from that same impact.

\subsection{Discussion on family results}
Given that the meteorite record suggests there existed $\sim$100 differentiated parent bodies, it has been surprising no evidence for significant differentiation of asteroids and within asteroid families has been seen. The photometric colors and albedos of the $\sim$100 known asteroid families \citep{Nesvorny2015, Milani2014} tend to be very homogeneous \citep{Parker2008, Masiero2011}. Through spectroscopic investigation, however, a number of families have been identified as remnants of differentiated parent bodies: Vesta \citep{Mccord1970,Consolmagno1977,Binzel1993}, Merxia, Agnia \citep{Sunshine2004, Vernazza2014}, Maria \citep{Fieber2011} and Hungaria \citep{Gaffey1992, Kelly2002}. Two families have been identified as candidates for at least partial differentiation: Eos and Eunomia. The Eos family was found to have mineral compositions consistent with forsteritic olivine \citep{MotheDiniz2005, MotheDiniz2008}. Other studies, however, have shown Eos more closely resembles CO and CV carbonaceous chondrite meteorites \citep{Bell1988,Doressoundiram1998,Clark2009}. The Eunomia family has been linked to partial differentiation by \citet{Nathues2005,Nathues2010}. For a review of the physical properties of asteroid families see \citet{Masiero2015}

In this work, we find very little evidence for full differentiation in families in the classical sense of forming a basaltic crust, olivine-rich mantle, and iron core. There were very few olivine-rich candidates available to survey to begin with, and most of those candidates were found to be false positives. The Merxia and Agnia families, for example, that are suggested to be candidates for differentiation, had 64 and 91 members with SDSS colors, respectively, with 0 and 6 differentiated candidates in each. None of those 6 candidates were observable in our survey. The Flora family is the only one with multiple ``mantle'' fragments, but their presence within the Flora family is not statistically distinct from the background A-type population.

\subsection{Implications for formation and evolution of small bodies and the solar system}
There are four broad explanations for why we do not see an abundance olivine-dominated (or basalt-rich) bodies within the main belt: These bodies 1. were ground down to sizes below our observational threshold in the ``battered to bits'' theory \citep{Burbine1996}, 2. are masked as another spectral type due to surface processes such as space weathering. 3. formed early in the terrestrial planet region and fragments were later implanted into the Main Belt, or 4. our understanding of asteroid differentiation is incomplete and a thick olivine-rich mantle may not be commonly formed.

We have established that there is no significant unknown mantle material in the main asteroid belt down to diameters of $\sim$2 km. We cannot yet systematically probe the compositions of sub-2 km main-belt asteroids to fully test the ``battered to bits'' scenario by \citet{Burbine1996}, but there are at least three additional constraints. First, simulations of Main-Belt collisional history in scenarios with more Vesta-like objects show that the currently observed Main Belt could not have had more than 3-4 Vesta-like objects, otherwise we would see more V-type asteroids not associated with Vesta \citep{Scott2015}. Second, if there were a significant population of sub-2 km A-type asteroids, their small sizes would cause them to be delivered relatively quickly to resonances and into planetary-crossing orbits. Either those bodies existed and were cleared out long ago with only a depleted population surviving, or we should see them in the NEO population. Through the fossil meteorite record it has been shown that the compositional makeup of the meteorite flux on Earth (and by proxy the NEO population from which they originate) has changed dramatically over time and is often dominated by specific events, particularly asteroid disruptions, that occur in the Main Belt \citep{Schmitz1997,Schmitz2003,Heck2004,Heck2017}. Near-infrared spectral surveys of NEOs do not find evidence for an abundance of A-types. Third, an explanation would be needed for why olivine-dominated bodies experienced more collisional erosion than the asteroid classes most abundant in the Main Belt today: the S- and C-types.

There are a number of processes that can affect the surfaces of asteroids, and thus change their interpreted composition from spectral measurements. Space weathering is one of the more common processes affecting the surfaces of airless bodies. Bombardment by high-energy particles and micrometeorites causes chemical changes to the surface material that produces a variety of changes in the measured properties such as changes in albedo as well as spectral band depth and spectral slope \citep[for a review, see][]{Clark2002, Chapman2004, Brunetto2015}. The continuum of the spectrum of olivine is neutral, whereas the A-type asteroids we measure have some of the reddest slopes in the inner solar system. This reddening effect has been reproduced in irradiation laboratory experiments that mimic the space environment \citep{Sasaki2001,Brunetto2006}. While the prominent one-micron absorption band becomes less pronounced after irradiation, it remains clearly identifiable. Shock darkening from collision is another process that affects the surfaces of asteroids \citep{Britt1989,Reddy2014,Kohout2014}. While shocking has been seen to darken the surfaces of ordinary chondrite meteorites to the point where the absorption band is severely depressed, it is not clear that this is a common process is in the asteroid belt or that it would act globally across multi-kilometer bodies. The fact that significant spectral diversity exists in the main belt and that collisions for multi-kilometer bodies are relatively rare suggests that while this is a notable process, it does not affect the majority of asteroids. Given these arguments, it is unlikely that a large population of olivine-dominated material is spectrally hidden within another class of objects.

The third explanation for the dearth of A-types is that differentiated planetesimals did not form in the Main Belt. Instead, they formed early in the terrestrial planet region at a point where the abundance of $^{26}$Al was high. Thermal modeling of some iron meteorite parent bodies that were heated by $^{26}$Al finds that these objects could have accreted less than 0.4 million years after the formation of CAIs (Calcium Aluminum Inclusions) \citep{Kruijer2017}, the earliest condensates in the Solar nebula.  In this tumultuous early period in Solar System history, collisions were much more frequent. Mantles could be stripped from their cores in events such as low-velocity hit-and-run collisions \citep{Asphaug2014,Scott2015}. Any fragments that survive these collisions would then need to be implanted into the Main Belt to remain on stable orbits to be observed today. Three dynamical scenarios have been proposed to achieve implantation: 1. gravitational scattering among planetary embryos \citep{Bottke2006}, 2. the Grand Tack planetary migration period \citep{Walsh2011,Walsh2012}, and 3. the Nice Model planetary migration period \citep{Morbidelli2005, Morbidelli2015, Scott2015}. See the discussion by \citet{Scott2015} in \textit{Asteroids IV} as dynamical implantation is their preferred scenario. Each of these methods may produce a different implantation signature, and our constraint of a uniform distribution may help distinguish among dynamical models.

Finally, it is also possible that iron meteorite parent bodies either did not form extensive olivine mantles or differentiation is hidden by a primitive crust that is still preserved. \citet{ElkinsTanton2011} modeled that partial differentiation of chondritic material could create a differentiated interior with an unheated crust. Asteroid (21) Lutetia is a candidate for this scenario because it has surface spectral properties consistent with a chondritic composition, but it has a high bulk density measured by the Rosetta mission. Work by \citet{Weiss2012} attribute this to internal differentiation, although \citet{Vernazza2011} suggest Lutetia's spectral and physical properties are consistent with the primitive meteorite class enstatite chondrites. Given the spectral homogeneity of asteroid families that essentially allow us to probe the interior of larger parent bodies, it is not clear that this type of partial differentiation is common in the asteroid belt. Dawn observations of (4) Vesta have found no evidence for an olivine-dominated mantle.  Large impact craters on Vesta's South Pole, which could have excavated down to depths of 60-100 km \citep{Clenet2014}, did not expose significant amounts of olivine \citep{Ammannito2013}.   \citet{Clenet2014} finds that the crust-mantle boundary is deeper than 80 km.  The range of Mg compositions of olivine in howardites \citep{Lunning2015}, which are believed to be fragments of Vesta,  are consistent with forming through partial melting \citep[e.g.,][]{Wilson2012} and not with full-scale melting in a whole-mantle magma ocean.  

A recent model of differentiation of Vesta-sized bodies argues that olivine-dominated mantles would not form \citep{ElkinsTanton2014}.  These models finds that the first crystallizing mineral would be olivine, which would settle to the core-mantle boundary.  Samples of these core-mantle boundaries would be pallasites.  However, due to the high viscosity of the molten mantle, only the earliest-forming crystals with sizes of several cm would settle.  The remaining mantle material would then solidify in bulk and not result in the significant accumulation of olivine crystals \citep{Scheinberg2015}.

\subsection{Future surveys to further this work}

The large survey data publicly available today such as SDSS  \citep{York2000,Ivezic2001}, WISE \citep{Mainzer2011,Masiero2011}, and VISTA \citep{McMahon2013,Popescu2016} with tens to hundreds of thousands observations have greatly advanced our ability to characterize the main-belt population as a whole and to focus in on finer details of the compositional structure of the main belt. We expect upcoming surveys will provide equally if not even greater insight through larger numbers of observations and better quality data. These include Gaia \citep{Gaia2016}, Euclid \citep{carry2018}, and  LSST \citep{LSST2009}. For questions related to olivine-dominated material and differentiation, we will look to these surveys to identify more candidates at broader locations and smaller sizes than previously available. 

By providing orders of magnitude improvement on the accuracy of orbits, and visible spectra for 300,000 asteroids \citep{mignard2007} Gaia will allow minute identification of family members, helping identifying interloper and opening the door to extensive searches of in-homogeneity in parent bodies of families.

The LSST is expected to discover millions of main-belt asteroids, and hundreds of thousands of NEAs. With multi-band photometry in a set of filter similar to that of the SDSS, broad compositional classification will be possible, pushing our knowledge to smaller sizes, and allowing a clear link to be established between the NEA and their source regions in the MB. The apparent contradiction of small samples of NEA as mentioned here should vanish in the LSST era.

However, both surveys will operate in the visible only. As shown here, the ambiguity in spectral classification will affect both, and near-infrared data will be crucial to identify rare mineralogies such as the olivine-rich A-types discussed in the present study. The ESA mission Euclid, to be launched in 2020, is expected to observe about 150,000 small bodies (mainly MB) over its six-year mission, in the visible and three infrared broad-band filters, enabling additional compositional investigations \citep{carry2018}.

\section{Conclusion}

The main findings and conclusions of this work are
\begin{itemize}
\item We confirm 21 A-type asteroids distributed throughout the main belt including the discovery of A-type asteroid (11616) in the Cybele region. We find 0.16\% of the main belt is A-type, a very small fraction. We also reaffirm that there is no significant undiscovered olivine-dominated material down to $\sim$2 km diameters. 
\item The fraction of A-types in the main belt is strikingly similar to what is found in the NEO population in the survey by \citep{Binzel2018} which ranges from 0.1-0.6\% depending on the data included.
\item We estimate the total number of A-types in the main belt down to $\sim$2 km to be $\sim$600. Most of the A-type mass ($>$ 80\%) is contained within the largest body (354) Eleonora. About half of A-types by number are within the smallest size bin (16$\leq$H$<$17).
\item The distribution as a function of semi-major axis is relatively flat. This flat distribution does not support a locally-formed theory as they span a wide semi-major axis range across the belt where the dominant type transitions dramatically from S to C to P. Instead, the results are more in line with the theory that these fragments were later implanted.
\item While we find 6 A- and V-type bodies dynamically associated with families, we find no statistically significant evidence that there has been differentiation in these families, at least in the canonical sense of forming a basaltic crust, olivine-rich mantle, and iron-rich core. This work supports evidence that asteroids in the main belt are generally not differentiated and that differentiated material did not form locally within the main belt.
\end{itemize}

\section*{Acknowledgments}
  Observations reported here were obtained at the NASA Infrared Telescope
  Facility, which is operated by the University of Hawaii under Cooperative
  Agreement NCC 5-538 with the National Aeronautics and Space Administration,
  Science Mission Directorate, Planetary Astronomy Program.  
  This paper includes data gathered with the 6.5 meter Magellan Telescopes
  located at Las Campanas Observatory, Chile.
  We acknowledge support from the Faculty of the European Space Astronomy 
  Centre (ESAC) for FD's visit. DP is grateful to the AXA research fund.
  This material is based upon work supported by the National Aeronautics and
  Space Administration under Grant No. NNX12AL26G issued through the Planetary
  Astronomy Program and by Hubble Fellowship grant HST-HF-51319.01-A awarded 
  by the Space Telescope Science Institute, which is operated by the Association of 
  Universities for Research in Astronomy, Inc., for NASA, under contract NAS 5-26555. 
  MIT researchers performing this work were supported by NASA grant 09-NEOO009-0001, 
  and by the National Science Foundation under Grants Nos. 0506716 and 0907766.  
HHH acknowledges support from NASA Solar System Workings program 80NSSC17K0723.  
THB would like to thank the Remote, In Situ, and Synchrotron Studies for Science and 
  Exploration (RIS4E) Solar System Exploration Research Virtual Institute (SSERVI) for support.
  Any opinions, findings, and conclusions or recommendations
  expressed in this article are those of the authors and do not necessarily
  reflect the views of the National Aeronautics and Space Administration.

\clearpage
\bibliographystyle{icarus.bst}

\begin{thebibliography}{}

\bibitem[{Ammannito} et~al.(2013){Ammannito}, {de Sanctis}, {Palomba},
  {Longobardo}, {Mittlefehldt}, {McSween}, {Marchi}, {Capria}, {Capaccioni},
  {Frigeri}, {Pieters}, {Ruesch}, {Tosi}, {Zambon}, {Carraro}, {Fonte},
  {Hiesinger}, {Magni}, {McFadden}, {Raymond}, {Russell}, and
  {Sunshine}]{Ammannito2013}
{Ammannito}, E., {de Sanctis}, M.~C., {Palomba}, E., {Longobardo}, A.,
  {Mittlefehldt}, D.~W., {McSween}, H.~Y., {Marchi}, S., {Capria}, M.~T.,
  {Capaccioni}, F., {Frigeri}, A., {Pieters}, C.~M., {Ruesch}, O., {Tosi}, F.,
  {Zambon}, F., {Carraro}, F., {Fonte}, S., {Hiesinger}, H., {Magni}, G.,
  {McFadden}, L.~A., {Raymond}, C.~A., {Russell}, C.~T., {Sunshine}, J.~M.,
  2013.
\newblock {Olivine in an unexpected location on Vesta's surface}.
\newblock \nat~504, 122--125.

\bibitem[{Asphaug} and {Reufer}(2014){Asphaug} and {Reufer}]{Asphaug2014}
{Asphaug}, E., {Reufer}, A., 2014.
\newblock {Mercury and other iron-rich planetary bodies as relics of
  inefficient accretion}.
\newblock Nature Geoscience~7, 564--568.

\bibitem[{Bell}(1988){Bell}]{Bell1988}
{Bell}, J.~F., 1988.
\newblock {A probable asteroidal parent body for the CO or CV chondrites}.
\newblock Meteoritics~23, 256--257.

\bibitem[{Bell} et~al.(1989){Bell}, {Davis}, {Hartmann}, and
  {Gaffey}]{Bell1989}
{Bell}, J.~F., {Davis}, D.~R., {Hartmann}, W.~K., {Gaffey}, M.~J., 1989.
\newblock {\em Asteroids - The Big Picture}.

\bibitem[{Benedix} et~al.(2014){Benedix}, {Haack}, and {McCoy}]{Benedix2014}
{Benedix}, G.~K., {Haack}, H., {McCoy}, T.~J., 2014.
\newblock {\em {Iron and Stony-Iron Meteorites}}, pp.\  267--285.

\bibitem[{Binzel} and {DeMeo}(tted){Binzel} and {DeMeo}]{Binzel2018}
{Binzel}, R.~P., {DeMeo}, F.~E. e.~a., 2018, submitted.
\newblock Compositional distributions and evolutionary processes for the
  near-earth object population: Results from the mit-hawaii near-earth object
  spectroscopic survey (mithneos).
\newblock Icarus.

\bibitem[{Binzel} et~al.(2004){Binzel}, {Rivkin}, {Stuart}, {Harris}, {Bus},
  and {Burbine}]{Binzel2004}
{Binzel}, R.~P., {Rivkin}, A.~S., {Stuart}, J.~S., {Harris}, A.~W., {Bus},
  S.~J., {Burbine}, T.~H., 2004.
\newblock {Observed spectral properties of near-Earth objects: results for
  population distribution, source regions, and space weathering processes}.
\newblock Icarus~170, 259--294.

\bibitem[{Binzel} et~al.(2006){Binzel}, {Thomas}, {DeMeo}, {Tokunaga},
  {Rivkin}, and {Bus}]{Binzel2006}
{Binzel}, R.~P., {Thomas}, C.~A., {DeMeo}, F.~E., {Tokunaga}, A., {Rivkin},
  A.~S., {Bus}, S.~J., 2006.
\newblock {The MIT-Hawaii-IRTF Joint Campaign for NEO Spectral Reconnaissance}.
\newblock In: {Mackwell}, S., {Stansbery}, E. (Eds.), 37th Annual Lunar and
  Planetary Science Conference, Volume~37 of {\em Lunar and Planetary Inst.
  Technical Report}, pp.\  1491.

\bibitem[{Binzel} and {Xu}(1993){Binzel} and {Xu}]{Binzel1993}
{Binzel}, R.~P., {Xu}, S., 1993.
\newblock {Chips off of asteroid 4 Vesta - Evidence for the parent body of
  basaltic achondrite meteorites}.
\newblock Science~260, 186--191.

\bibitem[{Borisov} et~al.(2017){Borisov}, {Christou}, {Bagnulo}, {Cellino},
  {Kwiatkowski}, and {Dell'Oro}]{Borisov2017}
{Borisov}, G., {Christou}, A., {Bagnulo}, S., {Cellino}, A., {Kwiatkowski}, T.,
  {Dell'Oro}, A., 2017.
\newblock {The olivine-dominated composition of the Eureka family of Mars
  Trojan asteroids}.
\newblock MNRAS~466, 489--495.

\bibitem[{Bottke} et~al.(2006){Bottke}, {Nesvorn{\'y}}, {Grimm}, {Morbidelli},
  and {O'Brien}]{Bottke2006}
{Bottke}, W.~F., {Nesvorn{\'y}}, D., {Grimm}, R.~E., {Morbidelli}, A.,
  {O'Brien}, D.~P., 2006.
\newblock {Iron meteorites as remnants of planetesimals formed in the
  terrestrial planet region}.
\newblock Nature~439, 821--824.

\bibitem[{Bottke} et~al.(2006){Bottke}, {Vokrouhlick{\'y}}, {Rubincam}, and
  {Nesvorn{\'y}}]{Bottke2006Yorp}
{Bottke}, W.~F., Jr., {Vokrouhlick{\'y}}, D., {Rubincam}, D.~P.,
  {Nesvorn{\'y}}, D., 2006.
\newblock {The Yarkovsky and Yorp Effects: Implications for Asteroid Dynamics}.
\newblock Annual Review of Earth and Planetary Sciences~34, 157--191.

\bibitem[{Britt} and {Pieters}(1989){Britt} and {Pieters}]{Britt1989}
{Britt}, D.~T., {Pieters}, C.~M., 1989.
\newblock {Black chondrite meteorites: An analysis of fall frequency and the
  distribution of petrologic types}.
\newblock Meteoritics~24, 255.

\bibitem[{Brunetto} et~al.(2015){Brunetto}, {Loeffler}, {Nesvorn{\'y}},
  {Sasaki}, and {Strazzulla}]{Brunetto2015}
{Brunetto}, R., {Loeffler}, M.~J., {Nesvorn{\'y}}, D., {Sasaki}, S.,
  {Strazzulla}, G., 2015.
\newblock {\em {Asteroid Surface Alteration by Space Weathering Processes}},
  pp.\  597--616.

\bibitem[{Brunetto} et~al.(2006){Brunetto}, {Romano}, {Blanco}, {Fonti},
  {Martino}, {Orofino}, and {Verrienti}]{Brunetto2006}
{Brunetto}, R., {Romano}, F., {Blanco}, A., {Fonti}, S., {Martino}, M.,
  {Orofino}, V., {Verrienti}, C., 2006.
\newblock {Space weathering of silicates simulated by nanosecond pulse UV
  excimer laser}.
\newblock Icarus~180, 546--554.

\bibitem[{Burbine}(2014){Burbine}]{Burbine2014}
{Burbine}, T.~H., 2014.
\newblock {\em {Asteroids. Treatise on Geochemistry (2nd Edition}}, Volume
  Volume 2: Planets, Asteroids, Comets and The Solar System, pp.\  365--415.

\bibitem[{Burbine} and {Binzel}(2002){Burbine} and {Binzel}]{Burbine2002}
{Burbine}, T.~H., {Binzel}, R.~P., 2002.
\newblock {Small Main-Belt Asteroid Spectroscopic Survey in the Near-Infrared}.
\newblock Icarus~159, 468--499.

\bibitem[{Burbine} et~al.(1996){Burbine}, {Meibom}, and {Binzel}]{Burbine1996}
{Burbine}, T.~H., {Meibom}, A., {Binzel}, R.~P., 1996.
\newblock {Mantle material in the main belt: Battered to bits?}
\newblock Meteoritics and Planetary Science~31, 607--620.

\bibitem[{Burns} et~al.(1979){Burns}, {Lamy}, and {Soter}]{Burns1979}
{Burns}, J.~A., {Lamy}, P.~L., {Soter}, S., 1979.
\newblock {Radiation forces on small particles in the solar system}.
\newblock Icarus~40, 1--48.

\bibitem[{Bus} and {Binzel}(2002){Bus} and {Binzel}]{Bus2002b}
{Bus}, S.~J., {Binzel}, R.~P., 2002.
\newblock {Phase II of the Small Main-Belt Asteroid Spectroscopic Survey, A
  Feature-Based Taxonomy}.
\newblock Icarus~158, 146--177.

\bibitem[{Carruba} et~al.(2005){Carruba}, {Michtchenko}, {Roig},
  {Ferraz-Mello}, and {Nesvorn{\'y}}]{carruba2005_vesta}
{Carruba}, V., {Michtchenko}, T.~A., {Roig}, F., {Ferraz-Mello}, S.,
  {Nesvorn{\'y}}, D., 2005.
\newblock {On the V-type asteroids outside the Vesta family. I. Interplay of
  nonlinear secular resonances and the Yarkovsky effect: the cases of 956 Elisa
  and 809 Lundia}.
\newblock \aap~441, 819--829.

\bibitem[{Carry}(2012){Carry}]{Carry2012}
{Carry}, B., 2012.
\newblock {Density of asteroids}.
\newblock Planetary and Space Science~73, 98--118.

\bibitem[{Carry}(2018){Carry}]{carry2018}
{Carry}, B., 2018.
\newblock {Solar system science with ESA Euclid}.
\newblock \aap~609, A113.

\bibitem[{Carvano} et~al.(2010){Carvano}, {Hasselmann}, {Lazzaro}, and
  {Moth{\'e}-Diniz}]{Carvano2010}
{Carvano}, J.~M., {Hasselmann}, P.~H., {Lazzaro}, D., {Moth{\'e}-Diniz}, T.,
  2010.
\newblock {SDSS-based taxonomic classification and orbital distribution of main
  belt asteroids}.
\newblock Astronomy and Astrophysics~510, A43.

\bibitem[{Cellino} et~al.(2002){Cellino}, {Bus}, {Doressoundiram}, and
  {Lazzaro}]{cellino2002_astfamilyspectroscopy_ast3}
{Cellino}, A., {Bus}, S.~J., {Doressoundiram}, A., {Lazzaro}, D., 2002.
\newblock {\em {Spectroscopic Properties of Asteroid Families}}, pp.\
  633--643.

\bibitem[{Chambers}(1999){Chambers}]{chambers1999_mercury}
{Chambers}, J.~E., 1999.
\newblock {A hybrid symplectic integrator that permits close encounters between
  massive bodies}.
\newblock \mnras~304, 793--799.

\bibitem[{Chapman}(1986){Chapman}]{Chapman1986}
{Chapman}, C.~R., 1986.
\newblock {Implications of the inferred compositions of asteroids for their
  collisional evolution}.
\newblock NASA and CNR, International Workshop on Catastrophic Disruption of
  Asteroids and Satellites~57, 103--114.

\bibitem[{Chapman}(2004){Chapman}]{Chapman2004}
{Chapman}, C.~R., 2004.
\newblock {Space Weathering of Asteroid Surfaces}.
\newblock Annual Review of Earth and Planetary Sciences~32, 539--567.

\bibitem[{Clark} et~al.(2002){Clark}, {Hapke}, {Pieters}, and
  {Britt}]{Clark2002}
{Clark}, B.~E., {Hapke}, B., {Pieters}, C., {Britt}, D., 2002.
\newblock {Asteroid Space Weathering and Regolith Evolution}.
\newblock Asteroids III, 585--599.

\bibitem[{Clark} et~al.(2009){Clark}, {Ockert-Bell}, {Cloutis}, {Nesvorny},
  {Moth{\'e}-Diniz}, and {Bus}]{Clark2009}
{Clark}, B.~E., {Ockert-Bell}, M.~E., {Cloutis}, E.~A., {Nesvorny}, D.,
  {Moth{\'e}-Diniz}, T., {Bus}, S.~J., 2009.
\newblock {Spectroscopy of K-complex asteroids: Parent bodies of carbonaceous
  meteorites?}
\newblock Icarus~202, 119--133.

\bibitem[{Clenet} et~al.(2014){Clenet}, {Jutzi}, {Barrat}, {Asphaug}, {Benz},
  and {Gillet}]{Clenet2014}
{Clenet}, H., {Jutzi}, M., {Barrat}, J.-A., {Asphaug}, E.~I., {Benz}, W.,
  {Gillet}, P., 2014.
\newblock {A deep crust-mantle boundary in the asteroid 4 Vesta}.
\newblock \nat~511, 303--306.

\bibitem[{Consolmagno} and {Drake}(1977){Consolmagno} and
  {Drake}]{Consolmagno1977}
{Consolmagno}, G.~J., {Drake}, M.~J., 1977.
\newblock {Composition and evolution of the eucrite parent body - Evidence from
  rare earth elements}.
\newblock Geochimica et Cosmochimica Acta~41, 1271--1282.

\bibitem[{Cruikshank} and {Hartmann}(1984){Cruikshank} and
  {Hartmann}]{Cruikshank1984}
{Cruikshank}, D.~P., {Hartmann}, W.~K., 1984.
\newblock {The meteorite-asteroid connection - Two olivine-rich asteroids}.
\newblock Science~223, 281--283.

\bibitem[{Cushing} et~al.(2004){Cushing}, {Vacca}, and {Rayner}]{Cushing2004}
{Cushing}, M.~C., {Vacca}, W.~D., {Rayner}, J.~T., 2004.
\newblock {Spextool: A Spectral Extraction Package for SpeX, a 0.8-5.5 Micron
  Cross-Dispersed Spectrograph}.
\newblock PASP~116, 362--376.

\bibitem[{de Le{\'o}n} et~al.(2004){de Le{\'o}n}, {Duffard}, {Licandro}, and
  {Lazzaro}]{deleon2004}
{de Le{\'o}n}, J., {Duffard}, R., {Licandro}, J., {Lazzaro}, D., 2004.
\newblock {Mineralogical characterization of A-type asteroid (1951) Lick}.
\newblock \aap~422, L59--L62.

\bibitem[{DeMeo} and {Binzel}(2008){DeMeo} and {Binzel}]{DeMeo2008}
{DeMeo}, F., {Binzel}, R.~P., 2008.
\newblock {Comets in the near-Earth object population}.
\newblock Icarus~194, 436--449.

\bibitem[{DeMeo}(2007){DeMeo}]{DeMeo2007}
{DeMeo}, F.~E., 2007.
\newblock Demeo taxonomy: Categorization of asteroids in the near-infrared.
\newblock Masters Thesis.

\bibitem[{DeMeo} et~al.(2015){DeMeo}, {Alexander}, {Walsh}, {Chapman}, and
  {Binzel}]{DeMeo2015}
{DeMeo}, F.~E., {Alexander}, C.~M.~O., {Walsh}, K.~J., {Chapman}, C.~R.,
  {Binzel}, R.~P., 2015.
\newblock {\em {The Compositional Structure of the Asteroid Belt}}, pp.\
  13--41.

\bibitem[{DeMeo} et~al.(2014){DeMeo}, {Binzel}, {Carry}, {Polishook}, and
  {Moskovitz}]{DeMeo2014D}
{DeMeo}, F.~E., {Binzel}, R.~P., {Carry}, B., {Polishook}, D., {Moskovitz},
  N.~A., 2014.
\newblock {Unexpected D-type interlopers in the inner main belt}.
\newblock Icarus~229, 392--399.

\bibitem[{DeMeo} et~al.(2009){DeMeo}, {Binzel}, {Slivan}, and
  {Bus}]{DeMeo2009taxo}
{DeMeo}, F.~E., {Binzel}, R.~P., {Slivan}, S.~M., {Bus}, S.~J., 2009.
\newblock {An extension of the Bus asteroid taxonomy into the near-infrared}.
\newblock Icarus~202, 160--180.

\bibitem[{DeMeo} and {Carry}(2013){DeMeo} and {Carry}]{DeMeo2013}
{DeMeo}, F.~E., {Carry}, B., 2013.
\newblock The taxonomic distribution of asteroids from multi-filter all-sky
  photometric surveys.
\newblock Icarus~226, 723--741.

\bibitem[{DeMeo} and {Carry}(2014){DeMeo} and {Carry}]{DeMeo2014}
{DeMeo}, F.~E., {Carry}, B., 2014.
\newblock {Solar System evolution from compositional mapping of the asteroid
  belt}.
\newblock Nature~505, 629--634.

\bibitem[{Doressoundiram} et~al.(1998){Doressoundiram}, {Barucci},
  {Fulchignoni}, and {Florczak}]{Doressoundiram1998}
{Doressoundiram}, A., {Barucci}, M.~A., {Fulchignoni}, M., {Florczak}, M.,
  1998.
\newblock {EOS Family: A Spectroscopic Study}.
\newblock Icarus~131, 15--31.

\bibitem[{Duffard}(2009){Duffard}]{Duffard2009}
{Duffard}, R., 2009.
\newblock {Basaltic Asteroids in the Solar System}.
\newblock Earth Moon and Planets~105, 221--226.

\bibitem[{Dykhuis} et~al.(2014){Dykhuis}, {Molnar}, {Van Kooten}, and
  {Greenberg}]{dykhuis2014_flora}
{Dykhuis}, M.~J., {Molnar}, L., {Van Kooten}, S.~J., {Greenberg}, R., 2014.
\newblock {Defining the Flora Family: Orbital properties, reflectance
  properties and age}.
\newblock \icarus~243, 111--128.

\bibitem[{Elkins-Tanton} et~al.(2014){Elkins-Tanton}, {Mandler}, and
  {Fu}]{ElkinsTanton2014}
{Elkins-Tanton}, L.~T., {Mandler}, B.~E., {Fu}, R.~R., 2014.
\newblock {Placing Vesta in the Range of Planetesimal Differentiation Models}.
\newblock In: Vesta in the Light of Dawn: First Exploration of a Protoplanet in
  the Asteroid Belt, Volume 1773 of {\em LPI Contributions}, pp.\  2034.

\bibitem[{Elkins-Tanton} et~al.(2011){Elkins-Tanton}, {Weiss}, and
  {Zuber}]{ElkinsTanton2011}
{Elkins-Tanton}, L.~T., {Weiss}, B.~P., {Zuber}, M.~T., 2011.
\newblock {Chondrites as samples of differentiated planetesimals}.
\newblock Earth and Planetary Science Letters~305, 1--10.

\bibitem[{Fieber-Beyer} et~al.(2011){Fieber-Beyer}, {Gaffey}, {Kelley},
  {Reddy}, {Reynolds}, and {Hicks}]{Fieber2011}
{Fieber-Beyer}, S.~K., {Gaffey}, M.~J., {Kelley}, M.~S., {Reddy}, V.,
  {Reynolds}, C.~M., {Hicks}, T., 2011.
\newblock {The Maria asteroid family: Genetic relationships and a plausible
  source of mesosiderites near the 3:1 Kirkwood Gap}.
\newblock Icarus~213, 524--537.

\bibitem[{Gaffey} et~al.(2015){Gaffey}, {Reddy}, {Fieber-Beyer}, and
  {Cloutis}]{Gaffey2015}
{Gaffey}, M.~J., {Reddy}, V., {Fieber-Beyer}, S., {Cloutis}, E., 2015.
\newblock {Asteroid (354) Eleonora: Plucking an odd duck}.
\newblock Icarus~250, 623--638.

\bibitem[{Gaffey} et~al.(1992){Gaffey}, {Reed}, and {Kelley}]{Gaffey1992}
{Gaffey}, M.~J., {Reed}, K.~L., {Kelley}, M.~S., 1992.
\newblock {Relationship of E-type Apollo asteroid 3103 (1982 BB) to the
  enstatite achondrite meteorites and the Hungaria asteroids}.
\newblock Icarus~100, 95--109.

\bibitem[{Gaia Collaboration} et~al.(2016){Gaia Collaboration}, {Prusti}, {de
  Bruijne}, {Brown}, {Vallenari}, {Babusiaux}, {Bailer-Jones}, {Bastian},
  {Biermann}, {Evans}, and et~al.]{Gaia2016}
{Gaia Collaboration}, {Prusti}, T., {de Bruijne}, J.~H.~J., {Brown}, A.~G.~A.,
  {Vallenari}, A., {Babusiaux}, C., {Bailer-Jones}, C.~A.~L., {Bastian}, U.,
  {Biermann}, M., {Evans}, D.~W., et~al., 2016.
\newblock {The Gaia mission}.
\newblock \aap~595, A1.

\bibitem[{Gil-Hutton} and {Licandro}(2010){Gil-Hutton} and
  {Licandro}]{Gil-Hutton2010}
{Gil-Hutton}, R., {Licandro}, J., 2010.
\newblock {Taxonomy of asteroids in the Cybele region from the analysis of the
  Sloan Digital Sky Survey colors}.
\newblock \icarus~206, 729--734.

\bibitem[{Heck} et~al.(2004){Heck}, {Schmitz}, {Baur}, {Halliday}, and
  {Wieler}]{Heck2004}
{Heck}, P.~R., {Schmitz}, B., {Baur}, H., {Halliday}, A.~N., {Wieler}, R.,
  2004.
\newblock {Fast delivery of meteorites to Earth after a major asteroid
  collision}.
\newblock Nature~430, 323--325.

\bibitem[{Heck} et~al.(2017){Heck}, {Schmitz}, {Bottke}, {Rout}, {Kita},
  {Cronholm}, {Defouilloy}, {Dronov}, and {Terfelt}]{Heck2017}
{Heck}, P.~R., {Schmitz}, B., {Bottke}, W.~F., {Rout}, S.~S., {Kita}, N.~T.,
  {Cronholm}, A., {Defouilloy}, C., {Dronov}, A., {Terfelt}, F., 2017.
\newblock {Rare meteorites common in the Ordovician period}.
\newblock Nature Astronomy~1, 0035.

\bibitem[{Ivezi{\'c}} et~al.(2002){Ivezi{\'c}}, {Lupton}, {Juri{\'c}},
  {Tabachnik}, {Quinn}, {Gunn}, {Knapp}, {Rockosi}, and
  {Brinkmann}]{ivezic2002_astfamilies}
{Ivezi{\'c}}, {\v Z}., {Lupton}, R.~H., {Juri{\'c}}, M., {Tabachnik}, S.,
  {Quinn}, T., {Gunn}, J.~E., {Knapp}, G.~R., {Rockosi}, C.~M., {Brinkmann},
  J., 2002.
\newblock {Color Confirmation of Asteroid Families}.
\newblock \aj~124, 2943--2948.

\bibitem[{Ivezi{\'c}} et~al.(2001){Ivezi{\'c}}, {Tabachnik}, {Rafikov},
  {Lupton}, {Quinn}, {Hammergren}, {Eyer}, {Chu}, {Armstrong}, {Fan},
  {Finlator}, {Geballe}, {Gunn}, {Hennessy}, {Knapp}, {Leggett}, {Munn},
  {Pier}, {Rockosi}, {Schneider}, {Strauss}, {Yanny}, {Brinkmann}, {Csabai},
  {Hindsley}, {Kent}, {Lamb}, {Margon}, {McKay}, {Smith}, {Waddel}, {York}, and
  {the SDSS Collaboration}]{Ivezic2001}
{Ivezi{\'c}}, {\v Z}., {Tabachnik}, S., {Rafikov}, R., {Lupton}, R.~H.,
  {Quinn}, T., {Hammergren}, M., {Eyer}, L., {Chu}, J., {Armstrong}, J.~C.,
  {Fan}, X., {Finlator}, K., {Geballe}, T.~R., {Gunn}, J.~E., {Hennessy},
  G.~S., {Knapp}, G.~R., {Leggett}, S.~K., {Munn}, J.~A., {Pier}, J.~R.,
  {Rockosi}, C.~M., {Schneider}, D.~P., {Strauss}, M.~A., {Yanny}, B.,
  {Brinkmann}, J., {Csabai}, I., {Hindsley}, R.~B., {Kent}, S., {Lamb}, D.~Q.,
  {Margon}, B., {McKay}, T.~A., {Smith}, J.~A., {Waddel}, P., {York}, D.~G.,
  {the SDSS Collaboration}, 2001.
\newblock {Solar System Objects Observed in the Sloan Digital Sky Survey
  Commissioning Data}.
\newblock Astronomical Journal~122, 2749--2784.

\bibitem[{Kelley} and {Gaffey}(2002){Kelley} and {Gaffey}]{Kelly2002}
{Kelley}, M.~S., {Gaffey}, M.~J., 2002.
\newblock {High-albedo asteroid 434 Hungaria: Spectrum, composition and genetic
  connections}.
\newblock Meteoritics and Planetary Science~37, 1815--1827.

\bibitem[{Kohout} et~al.(2014){Kohout}, {Gritsevich}, {Grokhovsky}, {Yakovlev},
  {Haloda}, {Halodova}, {Michallik}, {Penttil{\"a}}, and
  {Muinonen}]{Kohout2014}
{Kohout}, T., {Gritsevich}, M., {Grokhovsky}, V.~I., {Yakovlev}, G.~A.,
  {Haloda}, J., {Halodova}, P., {Michallik}, R.~M., {Penttil{\"a}}, A.,
  {Muinonen}, K., 2014.
\newblock {Mineralogy, reflectance spectra, and physical properties of the
  Chelyabinsk LL5 chondrite - Insight into shock-induced changes in asteroid
  regoliths}.
\newblock \icarus~228, 78--85.

\bibitem[{Kruijer} et~al.(2017){Kruijer}, {Burkhardt}, {Budde}, and
  {Kleine}]{Kruijer2017}
{Kruijer}, T.~S., {Burkhardt}, C., {Budde}, G., {Kleine}, T., 2017.
\newblock {Age of Jupiter inferred from the distinct genetics and formation
  times of meteorites}.
\newblock Proceedings of the National Academy of Science~114, 6712--6716.

\bibitem[{Kuroda} et~al.(2014){Kuroda}, {Ishiguro}, {Takato}, {Hasegawa},
  {Abe}, {Tsuda}, {Sugita}, {Usui}, {Hattori}, {Iwata}, {Imanishi}, {Terada},
  {Choi}, {Watanabe}, and {Yoshikawa}]{Kuroda2014}
{Kuroda}, D., {Ishiguro}, M., {Takato}, N., {Hasegawa}, S., {Abe}, M., {Tsuda},
  Y., {Sugita}, S., {Usui}, F., {Hattori}, T., {Iwata}, I., {Imanishi}, M.,
  {Terada}, H., {Choi}, Y.-J., {Watanabe}, S.-i., {Yoshikawa}, M., 2014.
\newblock {Visible-wavelength spectroscopy of subkilometer-sized near-Earth
  asteroids with a low delta-v}.
\newblock \pasj~66, 51.

\bibitem[{Leith} et~al.(2017){Leith}, {Moskovitz}, {Mayne}, {DeMeo}, {Takir},
  {Burt}, {Binzel}, and {Pefkou}]{Leith2017}
{Leith}, T.~B., {Moskovitz}, N.~A., {Mayne}, R.~G., {DeMeo}, F.~E., {Takir},
  D., {Burt}, B.~J., {Binzel}, R.~P., {Pefkou}, D., 2017.
\newblock {The compositional diversity of non-Vesta basaltic asteroids}.
\newblock Icarus~295, 61--73.

\bibitem[{Lord}(1992){Lord}]{Lord1992}
{Lord}, S.~D., 1992.
\newblock A new software tool for computing earth's atmospheric transmission of
  near- and far-infrared radiation.
\newblock NASA Tech. Mem.~(103957).

\bibitem[{LSST Science Collaboration} et~al.(2009){LSST Science Collaboration},
  {Abell}, {Allison}, {Anderson}, {Andrew}, {Angel}, {Armus}, {Arnett},
  {Asztalos}, {Axelrod}, and et~al.]{LSST2009}
{LSST Science Collaboration}, {Abell}, P.~A., {Allison}, J., {Anderson}, S.~F.,
  {Andrew}, J.~R., {Angel}, J.~R.~P., {Armus}, L., {Arnett}, D., {Asztalos},
  S.~J., {Axelrod}, T.~S., et~al., 2009.
\newblock {LSST Science Book, Version 2.0}.
\newblock ArXiv e-prints.

\bibitem[{Lunning} et~al.(2015){Lunning}, {McSween}, {Tenner}, {Kita}, and
  {Bodnar}]{Lunning2015}
{Lunning}, N.~G., {McSween}, H.~Y., {Tenner}, T.~J., {Kita}, N.~T., {Bodnar},
  R.~J., 2015.
\newblock {Olivine and pyroxene from the mantle of asteroid 4 Vesta}.
\newblock Earth and Planetary Science Letters~418, 126--135.

\bibitem[{Mainzer} et~al.(2011){Mainzer}, {Bauer}, {Grav}, {Masiero}, {Cutri},
  {Dailey}, {Eisenhardt}, {McMillan}, {Wright}, {Walker}, {Jedicke}, {Spahr},
  {Tholen}, {Alles}, {Beck}, {Brandenburg}, {Conrow}, {Evans}, {Fowler},
  {Jarrett}, {Marsh}, {Masci}, {McCallon}, {Wheelock}, {Wittman}, {Wyatt},
  {DeBaun}, {Elliott}, {Elsbury}, {Gautier}, {Gomillion}, {Leisawitz},
  {Maleszewski}, {Micheli}, and {Wilkins}]{Mainzer2011}
{Mainzer}, A., {Bauer}, J., {Grav}, T., {Masiero}, J., {Cutri}, R.~M.,
  {Dailey}, J., {Eisenhardt}, P., {McMillan}, R.~S., {Wright}, E., {Walker},
  R., {Jedicke}, R., {Spahr}, T., {Tholen}, D., {Alles}, R., {Beck}, R.,
  {Brandenburg}, H., {Conrow}, T., {Evans}, T., {Fowler}, J., {Jarrett}, T.,
  {Marsh}, K., {Masci}, F., {McCallon}, H., {Wheelock}, S., {Wittman}, M.,
  {Wyatt}, P., {DeBaun}, E., {Elliott}, G., {Elsbury}, D., {Gautier}, T., IV,
  {Gomillion}, S., {Leisawitz}, D., {Maleszewski}, C., {Micheli}, M.,
  {Wilkins}, A., 2011.
\newblock {Preliminary Results from NEOWISE: An Enhancement to the Wide-field
  Infrared Survey Explorer for Solar System Science}.
\newblock \apj~731, 53.

\bibitem[{Masi} et~al.(2008){Masi}, {Foglia}, and {Binzel}]{Masi2008}
{Masi}, G., {Foglia}, S., {Binzel}, R.~P., 2008.
\newblock {Search and Confirmation of V-type Asteroids Beyond 2.5 AU Using
  Sloan Digital Sky Survey Colors}.
\newblock In: Asteroids, Comets, Meteors 2008, Volume 1405 of {\em LPI
  Contributions}, pp.\  8065.

\bibitem[{Masiero} et~al.(2015){Masiero}, {DeMeo}, {Kasuga}, and
  {Parker}]{Masiero2015}
{Masiero}, J.~R., {DeMeo}, F.~E., {Kasuga}, T., {Parker}, A.~H., 2015.
\newblock {\em {Asteroid Family Physical Properties}}, pp.\  323--340.

\bibitem[{Masiero} et~al.(2013){Masiero}, {Mainzer}, {Bauer}, {Grav}, {Nugent},
  and {Stevenson}]{masiero2013_astfams_neowise}
{Masiero}, J.~R., {Mainzer}, A.~K., {Bauer}, J.~M., {Grav}, T., {Nugent},
  C.~R., {Stevenson}, R., 2013.
\newblock {Asteroid Family Identification Using the Hierarchical Clustering
  Method and WISE/NEOWISE Physical Properties}.
\newblock \apj~770, 7.

\bibitem[Masiero et~al.(2011)Masiero, Mainzer, Grav, Bauer, Cutri, Dailey,
  Eisenhardt, McMillan, Spahr, Skrutskie, Tholen, Walker, Wright, DeBaun,
  Elsbury, Gautier, Gomillion, and Wilkins]{Masiero2011}
Masiero, J.~R., Mainzer, A.~K., Grav, T., Bauer, J.~M., Cutri, R.~M., Dailey,
  J., Eisenhardt, P.~R.~M., McMillan, R.~S., Spahr, T.~B., Skrutskie, M.~F.,
  Tholen, D., Walker, R.~G., Wright, E.~L., DeBaun, E., Elsbury, D., Gautier,
  T., IV, Gomillion, S., Wilkins, A., 2011.
\newblock {Main Belt Asteroids with WISE/NEOWISE. I. Preliminary Albedos and
  Diameters}.
\newblock Astrophysical Journal~741, 68.

\bibitem[{Masiero} et~al.(2012){Masiero}, {Mainzer}, {Grav}, {Bauer}, {Cutri},
  {Nugent}, and {Cabrera}]{Masiero2012}
{Masiero}, J.~R., {Mainzer}, A.~K., {Grav}, T., {Bauer}, J.~M., {Cutri}, R.~M.,
  {Nugent}, C., {Cabrera}, M.~S., 2012.
\newblock {Preliminary Analysis of WISE/NEOWISE 3-Band Cryogenic and
  Post-cryogenic Observations of Main Belt Asteroids}.
\newblock \apjl~759, L8.

\bibitem[{McCord} et~al.(1970){McCord}, {Adams}, and {Johnson}]{Mccord1970}
{McCord}, T.~B., {Adams}, J.~B., {Johnson}, T.~V., 1970.
\newblock {Asteroid Vesta: Spectral Reflectivity and Compositional
  Implications}.
\newblock Science~168, 1445--1447.

\bibitem[{McMahon} et~al.(2013){McMahon}, {Banerji}, {Gonzalez}, {Koposov},
  {Bejar}, {Lodieu}, {Rebolo}, and {VHS Collaboration}]{McMahon2013}
{McMahon}, R.~G., {Banerji}, M., {Gonzalez}, E., {Koposov}, S.~E., {Bejar},
  V.~J., {Lodieu}, N., {Rebolo}, R., {VHS Collaboration}, 2013.
\newblock {First Scientific Results from the VISTA Hemisphere Survey (VHS)}.
\newblock The Messenger~154, 35--37.

\bibitem[{Migliorini} et~al.(1995){Migliorini}, {Zappal{\`a}}, {Vio}, and
  {Cellino}]{migliorini1995_interlopers}
{Migliorini}, F., {Zappal{\`a}}, V., {Vio}, R., {Cellino}, A., 1995.
\newblock {Interlopers within asteroid families.}
\newblock \icarus~118, 271--291.

\bibitem[{Mignard} et~al.(2007){Mignard}, {Cellino}, {Muinonen}, {Tanga},
  {Delb{\`o}}, {Dell'Oro}, {Granvik}, {Hestroffer}, {Mouret}, {Thuillot}, and
  {Virtanen}]{mignard2007}
{Mignard}, F., {Cellino}, A., {Muinonen}, K., {Tanga}, P., {Delb{\`o}}, M.,
  {Dell'Oro}, A., {Granvik}, M., {Hestroffer}, D., {Mouret}, S., {Thuillot},
  W., {Virtanen}, J., 2007.
\newblock {The Gaia Mission: Expected Applications to Asteroid Science}.
\newblock Earth Moon and Planets~101, 97--125.

\bibitem[{Milani} et~al.(2014){Milani}, {Cellino}, {Kne{\v z}evi{\'c}},
  {Novakovi{\'c}}, {Spoto}, and {Paolicchi}]{Milani2014}
{Milani}, A., {Cellino}, A., {Kne{\v z}evi{\'c}}, Z., {Novakovi{\'c}}, B.,
  {Spoto}, F., {Paolicchi}, P., 2014.
\newblock {Asteroid families classification: Exploiting very large datasets}.
\newblock Icarus~239, 46--73.

\bibitem[{Morbidelli} et~al.(2005){Morbidelli}, {Levison}, {Tsiganis}, and
  {Gomes}]{Morbidelli2005}
{Morbidelli}, A., {Levison}, H.~F., {Tsiganis}, K., {Gomes}, R., 2005.
\newblock {Chaotic capture of Jupiter's Trojan asteroids in the early Solar
  System}.
\newblock Nature~435, 462--465.

\bibitem[{Morbidelli} et~al.(2015){Morbidelli}, {Walsh}, {O'Brien}, {Minton},
  and {Bottke}]{Morbidelli2015}
{Morbidelli}, A., {Walsh}, K.~J., {O'Brien}, D.~P., {Minton}, D.~A., {Bottke},
  W.~F., 2015.
\newblock {\em {The Dynamical Evolution of the Asteroid Belt}}, pp.\  493--507.

\bibitem[{Moskovitz} et~al.(2008){Moskovitz}, {Jedicke}, {Gaidos}, {Willman},
  {Nesvorn{\'y}}, {Fevig}, and {Ivezi{\'c}}]{Moskovitz2008}
{Moskovitz}, N.~A., {Jedicke}, R., {Gaidos}, E., {Willman}, M., {Nesvorn{\'y}},
  D., {Fevig}, R., {Ivezi{\'c}}, {\v Z}., 2008.
\newblock {The distribution of basaltic asteroids in the Main Belt}.
\newblock Icarus~198, 77--90.

\bibitem[{Moth{\'e}-Diniz} and {Carvano}(2005){Moth{\'e}-Diniz} and
  {Carvano}]{MotheDiniz2005}
{Moth{\'e}-Diniz}, T., {Carvano}, J.~M., 2005.
\newblock {221 Eos: a remnant of a partially differentiated parent body?}
\newblock Astronomy and Astrophysics~442, 727--729.

\bibitem[{Moth{\'e}-Diniz} et~al.(2008){Moth{\'e}-Diniz}, {Carvano}, {Bus},
  {Duffard}, and {Burbine}]{MotheDiniz2008}
{Moth{\'e}-Diniz}, T., {Carvano}, J.~M., {Bus}, S.~J., {Duffard}, R.,
  {Burbine}, T.~H., 2008.
\newblock {Mineralogical analysis of the Eos family from near-infrared
  spectra}.
\newblock Icarus~195, 277--294.

\bibitem[{Nathues}(2010){Nathues}]{Nathues2010}
{Nathues}, A., 2010.
\newblock {Spectral study of the Eunomia asteroid family Part II: The small
  bodies}.
\newblock Icarus~208, 252--275.

\bibitem[{Nathues} et~al.(2005){Nathues}, {Mottola}, {Kaasalainen}, and
  {Neukum}]{Nathues2005}
{Nathues}, A., {Mottola}, S., {Kaasalainen}, M., {Neukum}, G., 2005.
\newblock {Spectral study of the Eunomia asteroid family. I. Eunomia}.
\newblock Icarus~175, 452--463.

\bibitem[{Nesvorn{\'y}}(2010){Nesvorn{\'y}}]{Nesvorny2010}
{Nesvorn{\'y}}, D., 2010.
\newblock {Nesvorny HCM Asteroid Families V1.0}.
\newblock NASA Planetary Data System~133,
  EAR--A--VARGBDET--5--NESVORNYFAM--V1.0.

\bibitem[{Nesvorny}(2015){Nesvorny}]{nesvorny2015_pdsastfam}
{Nesvorny}, D., 2015.
\newblock {Nesvorny HCM Asteroid Families V3.0}.
\newblock NASA Planetary Data System~EAR-A-VARGBDET-5-NESVORNYFAM-V3.0.

\bibitem[{Nesvorn{\'y}} et~al.(2015){Nesvorn{\'y}}, {Bro{\v z}}, and
  {Carruba}]{Nesvorny2015}
{Nesvorn{\'y}}, D., {Bro{\v z}}, M., {Carruba}, V., 2015.
\newblock {\em {Identification and Dynamical Properties of Asteroid Families}},
  pp.\  297--321.

\bibitem[{Nesvorn{\'y}} et~al.(2002){Nesvorn{\'y}}, {Morbidelli},
  {Vokrouhlick{\'y}}, {Bottke}, and {Bro{\v z}}]{nesvorny2002_flora}
{Nesvorn{\'y}}, D., {Morbidelli}, A., {Vokrouhlick{\'y}}, D., {Bottke}, W.~F.,
  {Bro{\v z}}, M., 2002.
\newblock {The Flora Family: A Case of the Dynamically Dispersed Collisional
  Swarm?}
\newblock \icarus~157, 155--172.

\bibitem[{Novakovi{\'c}} et~al.(2011){Novakovi{\'c}}, {Cellino}, and {Kne{\v
  z}evi{\'c}}]{novakovic2011_highifamilies}
{Novakovi{\'c}}, B., {Cellino}, A., {Kne{\v z}evi{\'c}}, Z., 2011.
\newblock {Families among high-inclination asteroids}.
\newblock \icarus~216, 69--81.

\bibitem[{Nugent} et~al.(2015){Nugent}, {Mainzer}, {Masiero}, {Bauer}, {Cutri},
  {Grav}, {Kramer}, {Sonnett}, {Stevenson}, and {Wright}]{Nugent2015}
{Nugent}, C.~R., {Mainzer}, A., {Masiero}, J., {Bauer}, J., {Cutri}, R.~M.,
  {Grav}, T., {Kramer}, E., {Sonnett}, S., {Stevenson}, R., {Wright}, E.~L.,
  2015.
\newblock {NEOWISE Reactivation Mission Year One: Preliminary Asteroid
  Diameters and Albedos}.
\newblock \apj~814, 117.

\bibitem[{Oszkiewicz} et~al.(2015){Oszkiewicz}, {Kankiewicz}, {W{\l}odarczyk},
  and {Kryszczy{\'n}ska}]{oszkiewicz2015_flora}
{Oszkiewicz}, D., {Kankiewicz}, P., {W{\l}odarczyk}, I., {Kryszczy{\'n}ska},
  A., 2015.
\newblock {Differentiation signatures in the Flora region}.
\newblock \aap~584, A18.

\bibitem[{Parker} et~al.(2008){Parker}, {Ivezi{\'c}}, {Juri{\'c}}, {Lupton},
  {Sekora}, and {Kowalski}]{Parker2008}
{Parker}, A., {Ivezi{\'c}}, {\v Z}., {Juri{\'c}}, M., {Lupton}, R., {Sekora},
  M.~D., {Kowalski}, A., 2008.
\newblock {The size distributions of asteroid families in the SDSS Moving
  Object Catalog 4}.
\newblock Icarus~198, 138--155.

\bibitem[{Perna} et~al.(2018){Perna}, {Barucci}, {Fulchignoni}, {Popescu},
  {Belskaya}, {Fornasier}, {Doressoundiram}, {Lantz}, and {Merlin}]{Perna2018}
{Perna}, D., {Barucci}, M.~A., {Fulchignoni}, M., {Popescu}, M., {Belskaya},
  I., {Fornasier}, S., {Doressoundiram}, A., {Lantz}, C., {Merlin}, F., 2018.
\newblock {A spectroscopic survey of the small near-Earth asteroid population:
  Peculiar taxonomic distribution and phase reddening}.
\newblock \planss~157, 82--95.

\bibitem[{Polishook} et~al.(2017){Polishook}, {Jacobson}, {Morbidelli}, and
  {Aharonson}]{Polishook2017}
{Polishook}, D., {Jacobson}, S.~A., {Morbidelli}, A., {Aharonson}, O., 2017.
\newblock {A Martian origin for the Mars Trojan asteroids}.
\newblock Nature Astronomy~1, 0179.

\bibitem[{Popescu} et~al.(2016){Popescu}, {Licandro}, {Morate}, {de Le{\'o}n},
  {Nedelcu}, {Rebolo}, {McMahon}, {Gonzalez-Solares}, and {Irwin}]{Popescu2016}
{Popescu}, M., {Licandro}, J., {Morate}, D., {de Le{\'o}n}, J., {Nedelcu},
  D.~A., {Rebolo}, R., {McMahon}, R.~G., {Gonzalez-Solares}, E., {Irwin}, M.,
  2016.
\newblock {Near-infrared colors of minor planets recovered from VISTA-VHS
  survey (MOVIS)}.
\newblock \aap~591, A115.

\bibitem[{Popescu} et~al.(2018){Popescu}, {Perna}, and {Barucci}]{Popescu2018}
{Popescu}, M., {Perna}, D., {Barucci}, M. A. e.~a., 2018.
\newblock Olivine-rich asteroids in the near-earth space.
\newblock MNRAS, submitted.

\bibitem[{Radovi{\'c}} et~al.(2017){Radovi{\'c}}, {Novakovi{\'c}}, {Carruba},
  and {Mar{\v c}eta}]{radovic2017_interlopers}
{Radovi{\'c}}, V., {Novakovi{\'c}}, B., {Carruba}, V., {Mar{\v c}eta}, D.,
  2017.
\newblock {An automatic approach to exclude interlopers from asteroid
  families}.
\newblock \mnras~470, 576--591.

\bibitem[{Rayner} et~al.(2003){Rayner}, {Toomey}, {Onaka}, {Denault},
  {Stahlberger}, {Vacca}, {Cushing}, and {Wang}]{Rayner2003}
{Rayner}, J.~T., {Toomey}, D.~W., {Onaka}, P.~M., {Denault}, A.~J.,
  {Stahlberger}, W.~E., {Vacca}, W.~E., {Cushing}, M.~C., {Wang}, S., 2003.
\newblock Spex: A medium-resolution 0.8-5.5 micron spectrograph and imager for
  the {NASA Infrared Telescope Facility}.
\newblock Astron. Soc. of the Pacific~115, 362--382.

\bibitem[{Reddy} et~al.(2014){Reddy}, {Sanchez}, {Bottke}, {Cloutis}, {Izawa},
  {O'Brien}, {Mann}, {Cuddy}, {Le Corre}, {Gaffey}, and {Fujihara}]{Reddy2014}
{Reddy}, V., {Sanchez}, J.~A., {Bottke}, W.~F., {Cloutis}, E.~A., {Izawa},
  M.~R.~M., {O'Brien}, D.~P., {Mann}, P., {Cuddy}, M., {Le Corre}, L.,
  {Gaffey}, M.~J., {Fujihara}, G., 2014.
\newblock {Chelyabinsk meteorite explains unusual spectral properties of
  Baptistina Asteroid Family}.
\newblock \icarus~237, 116--130.

\bibitem[{Rivkin}(2012){Rivkin}]{Rivkin2012}
{Rivkin}, A.~S., 2012.
\newblock {The fraction of hydrated C-complex asteroids in the asteroid belt
  from SDSS data}.
\newblock \icarus~221, 744--752.

\bibitem[{Rivkin} et~al.(2004){Rivkin}, {Binzel}, {Sunshine}, {Bus}, {Burbine},
  and {Saxena}]{Rivkin2004}
{Rivkin}, A.~S., {Binzel}, R.~P., {Sunshine}, J., {Bus}, S.~J., {Burbine},
  T.~H., {Saxena}, A., 2004.
\newblock {Infrared spectroscopic observations of 69230 Hermes (1937 UB):
  possible unweathered endmember among ordinary chondrite analogs}.
\newblock Icarus~172, 408--414.

\bibitem[{Rivkin} et~al.(2007){Rivkin}, {Trilling}, {Thomas}, {DeMeo}, {Spahr},
  and {Binzel}]{Rivkin2007}
{Rivkin}, A.~S., {Trilling}, D.~E., {Thomas}, C.~A., {DeMeo}, F., {Spahr},
  T.~B., {Binzel}, R.~P., 2007.
\newblock {Composition of the L5 Mars Trojans: Neighbors, not siblings}.
\newblock Icarus~192, 434--441.

\bibitem[{Sanchez} et~al.(2014){Sanchez}, {Reddy}, {Kelley}, {Cloutis},
  {Bottke}, {Nesvorn{\'y}}, {Lucas}, {Hardersen}, {Gaffey}, {Abell}, and
  {Corre}]{Sanchez2014}
{Sanchez}, J.~A., {Reddy}, V., {Kelley}, M.~S., {Cloutis}, E.~A., {Bottke},
  W.~F., {Nesvorn{\'y}}, D., {Lucas}, M.~P., {Hardersen}, P.~S., {Gaffey},
  M.~J., {Abell}, P.~A., {Corre}, L.~L., 2014.
\newblock {Olivine-dominated asteroids: Mineralogy and origin}.
\newblock Icarus~228, 288--300.

\bibitem[{Sasaki} et~al.(2001){Sasaki}, {Nakamura}, {Hamabe}, {Kurahashi}, and
  {Hiroi}]{Sasaki2001}
{Sasaki}, S., {Nakamura}, K., {Hamabe}, Y., {Kurahashi}, E., {Hiroi}, T., 2001.
\newblock {Production of iron nanoparticles by laser irradiation in a
  simulation of lunar-like space weathering}.
\newblock Nature~410, 555--557.

\bibitem[{Scheinberg} et~al.(2015){Scheinberg}, {Fu}, {Elkins-Tanton}, and
  {Weiss}]{Scheinberg2015}
{Scheinberg}, A., {Fu}, R.~R., {Elkins-Tanton}, L.~T., {Weiss}, B.~P., 2015.
\newblock {\em {Asteroid Differentiation: Melting and Large-Scale Structure}},
  pp.\  533--552.

\bibitem[{Schmitz} et~al.(2003){Schmitz}, {H{\"a}ggstr{\"o}m}, and
  {Tassinari}]{Schmitz2003}
{Schmitz}, B., {H{\"a}ggstr{\"o}m}, T., {Tassinari}, M., 2003.
\newblock {Sediment-Dispersed Extraterrestrial Chromite Traces a Major Asteroid
  Disruption Event}.
\newblock Science~300, 961--964.

\bibitem[{Schmitz} et~al.(1997){Schmitz}, {Peucker-Ehrenbrink},
  {Lindstr{\"o}m}, and {Tassinari}]{Schmitz1997}
{Schmitz}, B., {Peucker-Ehrenbrink}, B., {Lindstr{\"o}m}, M., {Tassinari}, M.,
  1997.
\newblock {Accretion rates of meteorites and cosmic dust in the Early
  Ordovician.}
\newblock Science~278, 88--90.

\bibitem[{Scott} et~al.(2015){Scott}, {Keil}, {Goldstein}, {Asphaug}, {Bottke},
  and {Moskovitz}]{Scott2015}
{Scott}, E.~R.~D., {Keil}, K., {Goldstein}, J.~I., {Asphaug}, E., {Bottke},
  W.~F., {Moskovitz}, N.~A., 2015.
\newblock {\em {Early Impact History and Dynamical Origin of Differentiated
  Meteorites and Asteroids}}, pp.\  573--595.

\bibitem[{Simcoe} et~al.(2013){Simcoe}, {Burgasser}, {Schechter}, {Fishner},
  {Bernstein}, {Bigelow}, {Pipher}, {Forrest}, {McMurtry}, {Smith}, and
  {Bochanski}]{Simcoe2013}
{Simcoe}, R.~A., {Burgasser}, A.~J., {Schechter}, P.~L., {Fishner}, J.,
  {Bernstein}, R.~A., {Bigelow}, B.~C., {Pipher}, J.~L., {Forrest}, W.,
  {McMurtry}, C., {Smith}, M.~J., {Bochanski}, J.~J., 2013.
\newblock {FIRE: A Facility Class Near-Infrared Echelle Spectrometer for the
  Magellan Telescopes}.
\newblock Publications of the Astronomical Society of the Pacific~125,
  270--286.

\bibitem[{Solontoi} et~al.(2012){Solontoi}, {Hammergren}, {Gyuk}, and
  {Puckett}]{Solontoi2012}
{Solontoi}, M.~R., {Hammergren}, M., {Gyuk}, G., {Puckett}, A., 2012.
\newblock {AVAST survey 0.4-1.0 {$\mu$}m spectroscopy of igneous asteroids in
  the inner and middle main belt}.
\newblock Icarus~220, 577--585.

\bibitem[{Sunshine} et~al.(2007){Sunshine}, {Bus}, {Corrigan}, {McCoy}, and
  {Burbine}]{Sunshine2007}
{Sunshine}, J.~M., {Bus}, S.~J., {Corrigan}, C.~M., {McCoy}, T.~J., {Burbine},
  T.~H., 2007.
\newblock {Olivine-dominated asteroids and meteorites: Distinguishing nebular
  and igneous histories}.
\newblock Meteoritics and Planetary Science~42, 155--170.

\bibitem[{Sunshine} et~al.(2004){Sunshine}, {Bus}, {McCoy}, {Burbine},
  {Corrigan}, and {Binzel}]{Sunshine2004}
{Sunshine}, J.~M., {Bus}, S.~J., {McCoy}, T.~J., {Burbine}, T.~H., {Corrigan},
  C.~M., {Binzel}, R.~P., 2004.
\newblock {High-calcium pyroxene as an indicator of igneous differentiation in
  asteroids and meteorites}.
\newblock Meteoritics and Planetary Science~39, 1343--1357.

\bibitem[{Tarduno} et~al.(2012){Tarduno}, {Cottrell}, {Nimmo}, {Hopkins},
  {Voronov}, {Erickson}, {Blackman}, {Scott}, and {McKinley}]{Tarduno2012}
{Tarduno}, J.~A., {Cottrell}, R.~D., {Nimmo}, F., {Hopkins}, J., {Voronov}, J.,
  {Erickson}, A., {Blackman}, E., {Scott}, E.~R.~D., {McKinley}, R., 2012.
\newblock {Evidence for a Dynamo in the Main Group Pallasite Parent Body}.
\newblock Science~338, 939.

\bibitem[{Tholen}(1984){Tholen}]{Tholen1984}
{Tholen}, D.~J., 1984.
\newblock {\em {Asteroid taxonomy from cluster analysis of photometry}}.
\newblock Ph.\ D. thesis, University of Arizona.

\bibitem[{Tody}(1993){Tody}]{Tody1993}
{Tody}, D., 1993.
\newblock {IRAF} in the nineties. in astronomical data.
\newblock In Astronomical Data Analysis Software and Systems II.

\bibitem[{Usui} et~al.(2011){Usui}, {Kuroda}, {M{\"u}ller}, {Hasegawa},
  {Ishiguro}, {Ootsubo}, {Ishihara}, {Kataza}, {Takita}, {Oyabu}, {Ueno},
  {Matsuhara}, and {Onaka}]{Usui2011}
{Usui}, F., {Kuroda}, D., {M{\"u}ller}, T.~G., {Hasegawa}, S., {Ishiguro}, M.,
  {Ootsubo}, T., {Ishihara}, D., {Kataza}, H., {Takita}, S., {Oyabu}, S.,
  {Ueno}, M., {Matsuhara}, H., {Onaka}, T., 2011.
\newblock {Asteroid Catalog Using Akari: AKARI/IRC Mid-Infrared Asteroid
  Survey}.
\newblock \pasj~63, 1117--1138.

\bibitem[{Vernazza} et~al.(2011){Vernazza}, {Lamy}, {Groussin}, {Hiroi},
  {Jorda}, {King}, {Izawa}, {Marchis}, {Birlan}, and {Brunetto}]{Vernazza2011}
{Vernazza}, P., {Lamy}, P., {Groussin}, O., {Hiroi}, T., {Jorda}, L., {King},
  P.~L., {Izawa}, M.~R.~M., {Marchis}, F., {Birlan}, M., {Brunetto}, R., 2011.
\newblock {Asteroid (21) Lutetia as a remnant of Earth's precursor
  planetesimals}.
\newblock \icarus~216, 650--659.

\bibitem[{Vernazza} et~al.(2014){Vernazza}, {Zanda}, {Binzel}, {Hiroi},
  {DeMeo}, {Birlan}, {Hewins}, {Ricci}, {Barge}, and {Lockhart}]{Vernazza2014}
{Vernazza}, P., {Zanda}, B., {Binzel}, R.~P., {Hiroi}, T., {DeMeo}, F.~E.,
  {Birlan}, M., {Hewins}, R., {Ricci}, L., {Barge}, P., {Lockhart}, M., 2014.
\newblock {Multiple and Fast: The Accretion of Ordinary Chondrite Parent
  Bodies}.
\newblock Astrophysical Journal~791, 120.

\bibitem[{Vokrouhlick{\'y}} et~al.(2017){Vokrouhlick{\'y}}, {Bottke}, and
  {Nesvorn{\'y}}]{vokrouhlicky2017_flora}
{Vokrouhlick{\'y}}, D., {Bottke}, W.~F., {Nesvorn{\'y}}, D., 2017.
\newblock {Forming the Flora Family: Implications for the Near-Earth Asteroid
  Population and Large Terrestrial Planet Impactors}.
\newblock \aj~153, 172.

\bibitem[{Walsh} et~al.(2011){Walsh}, {Morbidelli}, {Raymond}, {O'Brien}, and
  {Mandell}]{Walsh2011}
{Walsh}, K.~J., {Morbidelli}, A., {Raymond}, S.~N., {O'Brien}, D.~P.,
  {Mandell}, A.~M., 2011.
\newblock {A low mass for Mars from Jupiter's early gas-driven migration}.
\newblock Nature~475, 206--209.

\bibitem[{Walsh} et~al.(2012){Walsh}, {Morbidelli}, {Raymond}, {O'Brien}, and
  {Mandell}]{Walsh2012}
{Walsh}, K.~J., {Morbidelli}, A., {Raymond}, S.~N., {O'Brien}, D.~P.,
  {Mandell}, A.~M., 2012.
\newblock {Populating the asteroid belt from two parent source regions due to
  the migration of giant planets -``The Grand Tack''}.
\newblock Meteoritics and Planetary Science~47, 1941--1947.

\bibitem[{Weiss} et~al.(2012){Weiss}, {Elkins-Tanton}, {Antonietta Barucci},
  {Sierks}, {Snodgrass}, {Vincent}, {Marchi}, {Weissman}, {P{\"a}tzold},
  {Richter}, {Fulchignoni}, {Binzel}, and {Schulz}]{Weiss2012}
{Weiss}, B.~P., {Elkins-Tanton}, L.~T., {Antonietta Barucci}, M., {Sierks}, H.,
  {Snodgrass}, C., {Vincent}, J.-B., {Marchi}, S., {Weissman}, P.~R.,
  {P{\"a}tzold}, M., {Richter}, I., {Fulchignoni}, M., {Binzel}, R.~P.,
  {Schulz}, R., 2012.
\newblock {Possible evidence for partial differentiation of asteroid Lutetia
  from Rosetta}.
\newblock \planss~66, 137--146.

\bibitem[{Wilson} et~al.(2015){Wilson}, {Bland}, {Buczkowski}, {Keil}, and
  {Krot}]{Wilson2015}
{Wilson}, L., {Bland}, P.~A., {Buczkowski}, D., {Keil}, K., {Krot}, A.~N.,
  2015.
\newblock {\em {Hydrothermal and Magmatic Fluid Flow in Asteroids}}, pp.\
  553--572.

\bibitem[{Wilson} and {Keil}(2012){Wilson} and {Keil}]{Wilson2012}
{Wilson}, L., {Keil}, K., 2012.
\newblock {Volcanic activity on differentiated asteroids: A review and
  analysis}.
\newblock Chemie der Erde / Geochemistry~72, 289--321.

\bibitem[{Ye}(2011){Ye}]{Ye2011}
{Ye}, Q.-z., 2011.
\newblock {BVRI Photometry of 53 Unusual Asteroids}.
\newblock \aj~141, 32.

\bibitem[{York} et~al.(2000){York}, {Adelman}, {Anderson}, {Anderson}, {Annis},
  {Bahcall}, {Bakken}, {Barkhouser}, {Bastian}, {Berman}, {Boroski}, {Bracker},
  {Briegel}, {Briggs}, {Brinkmann}, {Brunner}, {Burles}, {Carey}, {Carr},
  {Castander}, {Chen}, {Colestock}, {Connolly}, {Crocker}, {Csabai},
  {Czarapata}, {Davis}, {Doi}, {Dombeck}, {Eisenstein}, {Ellman}, {Elms},
  {Evans}, {Fan}, {Federwitz}, {Fiscelli}, {Friedman}, {Frieman}, {Fukugita},
  {Gillespie}, {Gunn}, {Gurbani}, {de Haas}, {Haldeman}, {Harris}, {Hayes},
  {Heckman}, {Hennessy}, {Hindsley}, {Holm}, {Holmgren}, {Huang}, {Hull},
  {Husby}, {Ichikawa}, {Ichikawa}, {Ivezi{\'c}}, {Kent}, {Kim}, {Kinney},
  {Klaene}, {Kleinman}, {Kleinman}, {Knapp}, {Korienek}, {Kron}, {Kunszt},
  {Lamb}, {Lee}, {Leger}, {Limmongkol}, {Lindenmeyer}, {Long}, {Loomis},
  {Loveday}, {Lucinio}, {Lupton}, {MacKinnon}, {Mannery}, {Mantsch}, {Margon},
  {McGehee}, {McKay}, {Meiksin}, {Merelli}, {Monet}, {Munn}, {Narayanan},
  {Nash}, {Neilsen}, {Neswold}, {Newberg}, {Nichol}, {Nicinski}, {Nonino},
  {Okada}, {Okamura}, {Ostriker}, {Owen}, {Pauls}, {Peoples}, {Peterson},
  {Petravick}, {Pier}, {Pope}, {Pordes}, {Prosapio}, {Rechenmacher}, {Quinn},
  {Richards}, {Richmond}, {Rivetta}, {Rockosi}, {Ruthmansdorfer}, {Sandford},
  {Schlegel}, {Schneider}, {Sekiguchi}, {Sergey}, {Shimasaku}, {Siegmund},
  {Smee}, {Smith}, {Snedden}, {Stone}, {Stoughton}, {Strauss}, {Stubbs},
  {SubbaRao}, {Szalay}, {Szapudi}, {Szokoly}, {Thakar}, {Tremonti}, {Tucker},
  {Uomoto}, {Vanden Berk}, {Vogeley}, {Waddell}, {Wang}, {Watanabe},
  {Weinberg}, {Yanny}, {Yasuda}, and {SDSS Collaboration}]{York2000}
{York}, D.~G., {Adelman}, J., {Anderson}, J.~E., Jr., {Anderson}, S.~F.,
  {Annis}, J., {Bahcall}, N.~A., {Bakken}, J.~A., {Barkhouser}, R., {Bastian},
  S., {Berman}, E., {Boroski}, W.~N., {Bracker}, S., {Briegel}, C., {Briggs},
  J.~W., {Brinkmann}, J., {Brunner}, R., {Burles}, S., {Carey}, L., {Carr},
  M.~A., {Castander}, F.~J., {Chen}, B., {Colestock}, P.~L., {Connolly}, A.~J.,
  {Crocker}, J.~H., {Csabai}, I., {Czarapata}, P.~C., {Davis}, J.~E., {Doi},
  M., {Dombeck}, T., {Eisenstein}, D., {Ellman}, N., {Elms}, B.~R., {Evans},
  M.~L., {Fan}, X., {Federwitz}, G.~R., {Fiscelli}, L., {Friedman}, S.,
  {Frieman}, J.~A., {Fukugita}, M., {Gillespie}, B., {Gunn}, J.~E., {Gurbani},
  V.~K., {de Haas}, E., {Haldeman}, M., {Harris}, F.~H., {Hayes}, J.,
  {Heckman}, T.~M., {Hennessy}, G.~S., {Hindsley}, R.~B., {Holm}, S.,
  {Holmgren}, D.~J., {Huang}, C.-h., {Hull}, C., {Husby}, D., {Ichikawa},
  S.-I., {Ichikawa}, T., {Ivezi{\'c}}, {\v Z}., {Kent}, S., {Kim}, R.~S.~J.,
  {Kinney}, E., {Klaene}, M., {Kleinman}, A.~N., {Kleinman}, S., {Knapp},
  G.~R., {Korienek}, J., {Kron}, R.~G., {Kunszt}, P.~Z., {Lamb}, D.~Q., {Lee},
  B., {Leger}, R.~F., {Limmongkol}, S., {Lindenmeyer}, C., {Long}, D.~C.,
  {Loomis}, C., {Loveday}, J., {Lucinio}, R., {Lupton}, R.~H., {MacKinnon}, B.,
  {Mannery}, E.~J., {Mantsch}, P.~M., {Margon}, B., {McGehee}, P., {McKay},
  T.~A., {Meiksin}, A., {Merelli}, A., {Monet}, D.~G., {Munn}, J.~A.,
  {Narayanan}, V.~K., {Nash}, T., {Neilsen}, E., {Neswold}, R., {Newberg},
  H.~J., {Nichol}, R.~C., {Nicinski}, T., {Nonino}, M., {Okada}, N., {Okamura},
  S., {Ostriker}, J.~P., {Owen}, R., {Pauls}, A.~G., {Peoples}, J., {Peterson},
  R.~L., {Petravick}, D., {Pier}, J.~R., {Pope}, A., {Pordes}, R., {Prosapio},
  A., {Rechenmacher}, R., {Quinn}, T.~R., {Richards}, G.~T., {Richmond}, M.~W.,
  {Rivetta}, C.~H., {Rockosi}, C.~M., {Ruthmansdorfer}, K., {Sandford}, D.,
  {Schlegel}, D.~J., {Schneider}, D.~P., {Sekiguchi}, M., {Sergey}, G.,
  {Shimasaku}, K., {Siegmund}, W.~A., {Smee}, S., {Smith}, J.~A., {Snedden},
  S., {Stone}, R., {Stoughton}, C., {Strauss}, M.~A., {Stubbs}, C., {SubbaRao},
  M., {Szalay}, A.~S., {Szapudi}, I., {Szokoly}, G.~P., {Thakar}, A.~R.,
  {Tremonti}, C., {Tucker}, D.~L., {Uomoto}, A., {Vanden Berk}, D., {Vogeley},
  M.~S., {Waddell}, P., {Wang}, S.-i., {Watanabe}, M., {Weinberg}, D.~H.,
  {Yanny}, B., {Yasuda}, N., {SDSS Collaboration}, 2000.
\newblock {The Sloan Digital Sky Survey: Technical Summary}.
\newblock Astronomical Journal~120, 1579--1587.

\bibitem[{Zappala} et~al.(1990){Zappala}, {Cellino}, {Farinella}, and
  {Knezevic}]{zappala1990_hcm}
{Zappala}, V., {Cellino}, A., {Farinella}, P., {Knezevic}, Z., 1990.
\newblock {Asteroid families. I - Identification by hierarchical clustering and
  reliability assessment}.
\newblock \aj~100, 2030--2046.

\bibitem[{Zappala} et~al.(1994){Zappala}, {Cellino}, {Farinella}, and
  {Milani}]{zappala1994_hcm}
{Zappala}, V., {Cellino}, A., {Farinella}, P., {Milani}, A., 1994.
\newblock {Asteroid families. 2: Extension to unnumbered multiopposition
  asteroids}.
\newblock \aj~107, 772--801.

\end{thebibliography}

\end{document}